\documentclass[aps,prl,twocolumn,reprint, superscriptaddress]{revtex4-2}

\usepackage[caption=false]{subfig}
\usepackage{textgreek}
\usepackage{soul}
\usepackage{amsmath}
\usepackage{amssymb}
\usepackage{bm}
\usepackage{physics}
\usepackage{verbatim}
\usepackage[scr]{rsfso}
\usepackage{mwe}
\usepackage{graphicx}
\usepackage{overpic}
\usepackage[dvipsnames]{xcolor}
\definecolor{urlblue}{RGB}{6,69,173}

\usepackage[
	colorlinks=true,
	linkcolor=blue,
	anchorcolor=blue,
	citecolor=blue,
	urlcolor=blue,
	pdftitle={graviton_fci!}
]{hyperref}
\usepackage{tikz}
\usetikzlibrary{decorations.markings}

\tikzset{
  ->-/.style={
    decoration={markings, mark=at position #1 with {\arrow{latex}}},
    postaction={decorate}
  }
}

\newcommand{\ee}{\mathrm{e}}

\newcommand{\newsect}[1]{\noindent \textit{\textcolor{blue}{#1.--}}}

\makeatletter
\newcommand{\thetitle}{\@title}
\makeatother

\begin{document}

\title{Spectroscopy of phonon-coupled integer and fractional Chern insulators: emergence of polarons and chirality deficit of graviton mode}
\author{Min Long}
 \email{minlo@connect.hku.hk}
\affiliation{Department of Physics and HK Institute of Quantum Science \& Technology, The University of Hong Kong, Pokfulam Road,  Hong Kong SAR, China}
\affiliation{State Key Laboratory of Optical Quantum Materials, The University of Hong Kong, Pokfulam Road,  Hong Kong SAR, China}

\author{Yuzhu Wang}
\affiliation{Universi\'te Grenoble Alpes, Laboratoire de Physique et Mod\'elisation des Milieux Condens\'es (LPMMC),
Centre national de la recherche scientifique (CNRS), Grenoble, 38000 France}

% \author{Hongyu Lu}
% \affiliation{New Cornerstone Science Lab, Department of Physics, The University of Hong Kong, Pokfulam Road, Hong Kong SAR, China}
% \affiliation{HK Institute of Quantum Science \& Technology, The University of Hong Kong, Pokfulam Road, Hong Kong SAR, China}
% \affiliation{State Key Laboratory of Optical Quantum Materials, The University of Hong Kong, Pokfulam Road,  Hong Kong SAR, China}

 \author{Zi Yang Meng}
 \email{zymeng@hku.hk}
\affiliation{Department of Physics and HK Institute of Quantum Science \& Technology, The University of Hong Kong, Pokfulam Road,  Hong Kong SAR, China}
\affiliation{State Key Laboratory of Optical Quantum Materials, The University of Hong Kong, Pokfulam Road,  Hong Kong SAR, China}

\date{\today}

\begin{abstract} 
Strong electron-phonon coupling can destabilize both Chern insulators (CIs) and fractional Chern insulators (FCIs) in favor of charge order, but whether the instabilities in CI and FCI share a common microscopic mechanism remains unclear. In this paper, we address this question by studying a flat-band Haldane model coupled to dynamical Holstein phonons at integer and fractional fillings using density-matrix renormalization group with local basis optimization. We find that the low-energy effects of phonons are dominated by their dressing of charge-neutral collective modes. In the CI and FCI, phonons dress excitons and magnetorotons into composite modes that we identify as exciton polarons and magnetoroton polarons, respectively. These modes soften strongly upon approaching the transition, while the momenta of their energy minima anticipate the ordering wave vectors of the charge-ordered phases. The phonon spectrum, in turn, acquires dispersive features inherited from the exciton and magnetoroton modes, providing direct lattice signatures of these neutral excitations. Within the FCI phase, electron-phonon coupling also enhances the opposite-chirality spectral weight of the graviton response, thereby resulting in a chirality deficit. Our results provide a unified excitation-based picture of phonon-driven instabilities in CIs and FCIs and establish lattice dynamics as a probe of their neutral collective modes.
% Collective excitations encode the dynamics of correlated topological phases. Can lattice vibrations contribute to these excitations and characterize the phases that host them? Using density-matrix renormalization group with local basis optimization, we study the spinless flat-band Haldane model coupled to dynamical Holstein phonons at integer and fractional fillings. We find that electron–phonon coupling mainly dresses charge-neutral excitations and imprints their signatures on the phonon spectrum. In particular, phonons dress electron–hole excitons into exciton-polarons in the Chern insulator; in the fractional Chern insulator(FCI), they dress magnetorotons, charge-neutral bound states of fractionally charged quasiparticles and quasiholes, into magnetoroton-polarons,\minlo{Graviton stuffs}. At stronger coupling, charge order emerges in both regimes, including a magnetoroton-polaron-driven transition from the FCI to a charge-density wave state despite the electron density being dilute and the interaction being short-ranged. Our results establish lattice dynamics as a probe of neutral collective excitations and how electron–phonon coupling can destabilize a fractionalized topological phase toward charge order. \zymeng{(don't use unnecessary abbreviations.)}
\end{abstract}
%%%%%%
\maketitle
%%%%%%
\newsect{Introduction}
Charge-neutral modes are the characterising collective excitations of fractional quantum Hall (FQH) states. The pioneering work by Girvin, MacDonald, and
Platzman established the magnetoroton theory \cite{gmp1986magnetoroton} to describe the low-lying density collective excitations in FQH states. When the magnetoroton softens and its gap closes, the continuous translation symmetry will break, and the characteristic momentum of the magnetoroton sets the ordering wave vector of the emerging charge-density-wave (CDW) phase. Another example of charge-neutral excitations in FQH system is the Chiral graviton mode (CGM), a charge-neutral quadrupolar mode with angular momentum $L = 2$. The CGM describes the fluctuation of emergent metric degrees of freedom in an FQH droplet~\cite{haldane2011geometrical,wang2023Geometric} and has been observed recently~\cite{liang2024evidence}. Softening of the chiral graviton mode to zero energy can signal a transition to a fractional quantum Hall nematic phase, in which the ground state spontaneously breaks continuous rotational symmetry.~\cite{You2014TheoryNematic,pu2024microscopic,Bo2020Microscopic} 

The softening of neutral collective excitations can reveal instabilities of fractional quantum Hall phases and point to ways of tuning them. Recent advances in cavity quantum electrodynamics have shown that coupling electrons to long-wavelength cavity vacuum modes can modify the ground state~\cite{appugliese2022breakdown} and reshape chiral graviton excitations~\cite{bacciconi_prx2025_gravitonpolaritons}. By comparison, the effects of lattice phonons on neutral collective excitations remain largely unexplored. Studies of electron–phonon coupling in Chern insulators (CIs) and fractional Chern insulators (FCIs)~\cite{Tang2011_flat_chern_band, Sun2011_flat_chern_band, Neupert2011_flat_chern_band, Sheng2011_FQAH_checkerboard_fermion, Regnault2011_FCI,Xiao2011_quantum_hall,wuAdiabatic2012,KaiDasSarma_prl2011_flatbandsCB} have focused primarily on ground-state properties. In both CI and FCI, sufficiently strong coupling can destabilize the topological phase and favor charge order~\cite{Cangemi2019Topological,SousaJnior2026Realspace,Zezhu2026HaldaneHolstein}. This situation naturally prompts two questions: how do phonons modify the neutral excitations of FCIs, particularly in comparison with cavity modes? And does a common microscopic mechanism involving the coupling of
phonons to neutral excitations underlie the transitions to charge order
in CIs and FCIs?

The recent realization of fractional Chern insulators (FCIs), manifested by the fractional quantum anomalous Hall (FQAH) effect in two-dimensional moi\'re materials~\cite{Cai2023_signature_fqah_mote2,Park2023_observation_fqah_mote2,Zeng2023_thermo_evidence_fqah_mote2,Xu2023_Observation_FQAH_tMote2,Lu2024_FQAH_multilayer_graphene} has further motivated our study of FQH physics with lattice phonons. Unlike the predominantly long-wavelength vacuum modes of a cavity, phonons span a range of momenta and can therefore couple to neutral excitations throughout the Brillouin zone, including the magnetoroton at finite momentum. This motivates us to examine how electron–phonon coupling reshapes the neutral excitation spectrum of FCIs and whether the softening of these modes provides a microscopic route toward the charge-ordered phases found at strong coupling.

\begin{figure}
    \centering
    \begin{overpic}[width=\linewidth]{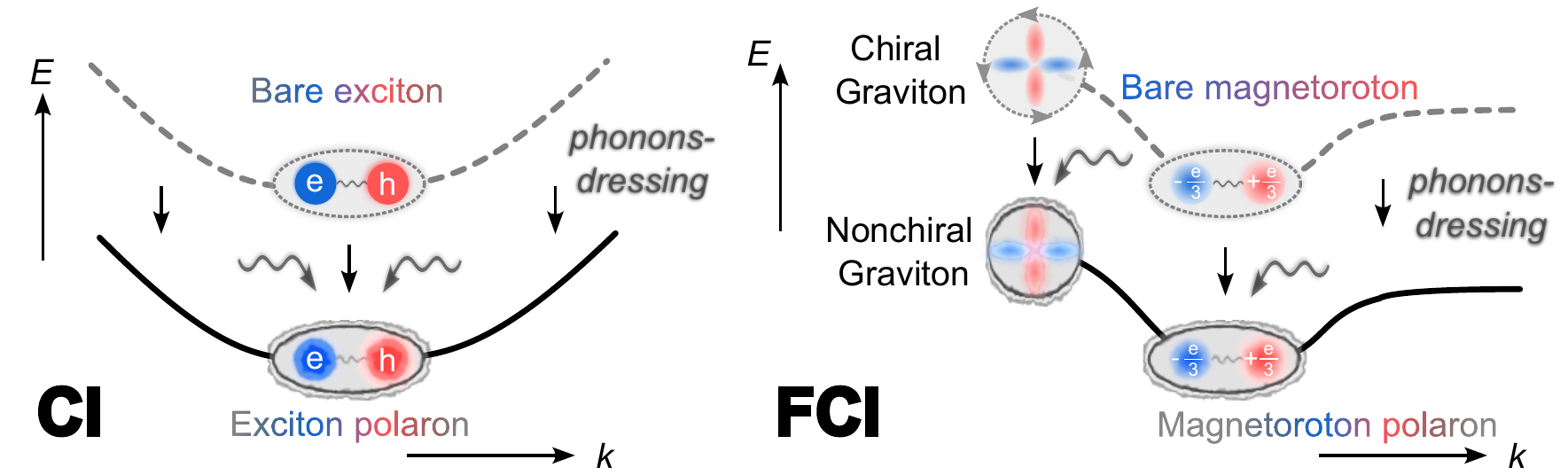}
    \put(-2,30){(a)}
    \put(50,30){(b)}
    \end{overpic}
    \par\vspace{5mm}
    \begin{overpic}[width=0.9\linewidth]{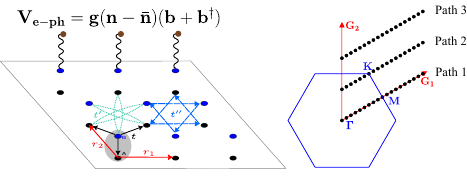}
    \put(-8,35){(c)}
    \put(51,35){(d)}
    \end{overpic}
   \caption{
\textbf{Phonon dressing of neutral excitations in CI and FCI, Hamiltonian and Brillouin zone.}
(a) In the CI, phonons dress excitons into an exciton polaron and soften its
dispersion. (b) In FCI, phonons dress the
magnetoroton, neutral bound state of fractionally charged
quasielectrons and quasiholes, into a magnetoroton polaron. Phonon
coupling also enhances spectral weight to the opposite-chirality
graviton response, producing a chirality deficit. The black and green
curves schematically represent the bare and phonon-dressed
dispersions, respectively. (c) Haldane--Holstein model on the
honeycomb lattice
(d) Brillouin Zone and the three momentum cuts accessible on the
$L_y=3$ cylinder.
}
    \label{fig:fig1}
\end{figure}

\begin{figure}
    \centering
    \begin{overpic}[width=\linewidth]
        {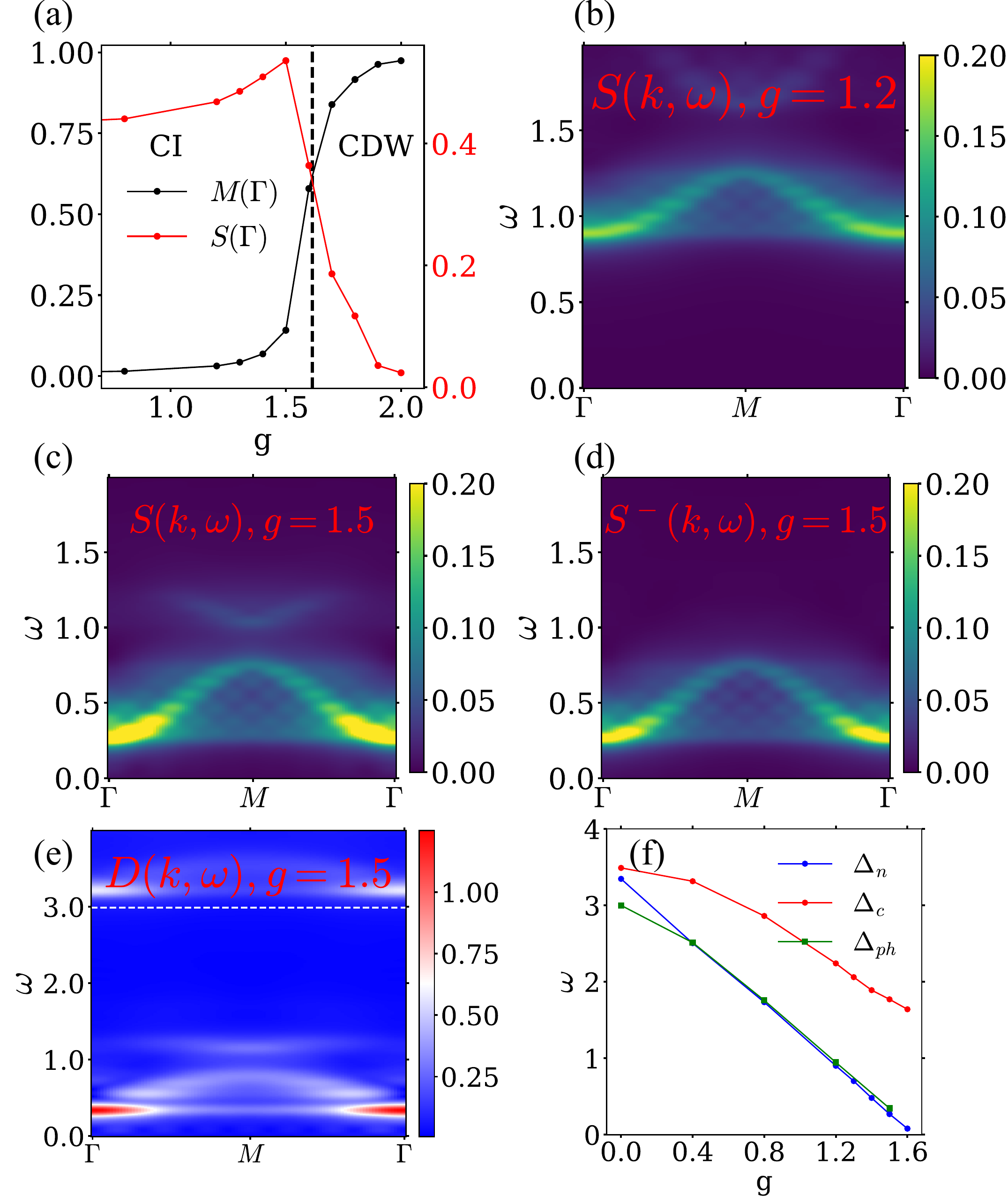}
        \put(2,91){(a)}
        \put(51,91){(b)}
    \end{overpic}

    \caption{\textbf{CI-CDW transition and Exciton-polarons }. The spectrum shown for the CI case focuses on path 1 in Fig.~\ref{fig:fig1} (d). (a) Evolution of CDW parameter $M(\Gamma)$ and static density correlation $S(\Gamma)$ at $\nu = 1$ filling as $g$ increases. (b) and (c) $S(\mathbf{k},\omega)$ defined in Eq.~\eqref{eq:spectra function} for different $g$, where the low-energy branch is fully captured by the $n_-$ branch as shown in (d). (e) Phonon spectral function $D(\mathbf{k},\omega)$; the white dashed line denotes the intrinsic frequency. (f)Evolution of neutral gap $\Delta_n$, charge gap $\Delta_c$, and phonon gap $\Delta_{ph}$ within the CI phase.}
    \label{fig:fig2}
\end{figure}

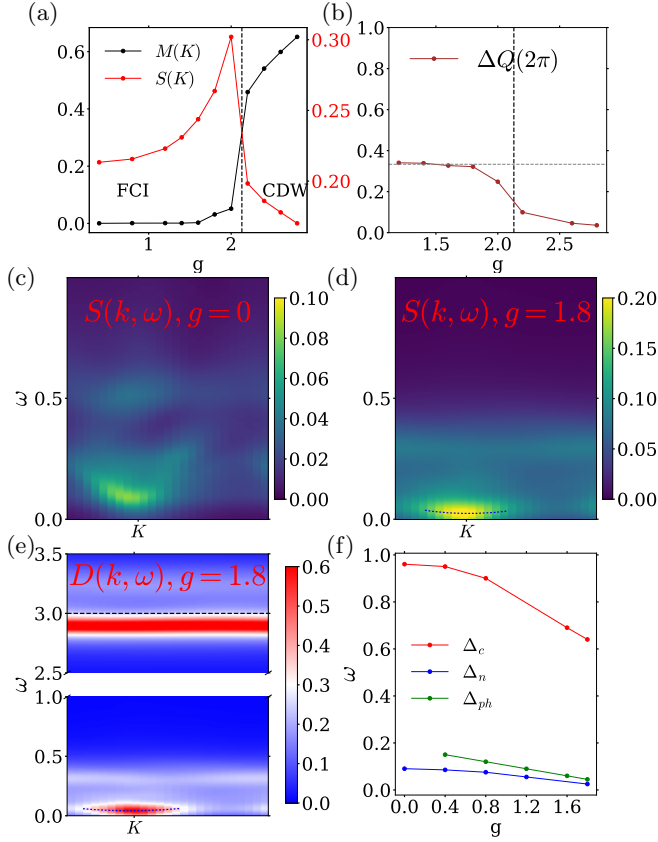
\begin{figure}
    \centering
    \begin{overpic}
        [width=0.85\linewidth]{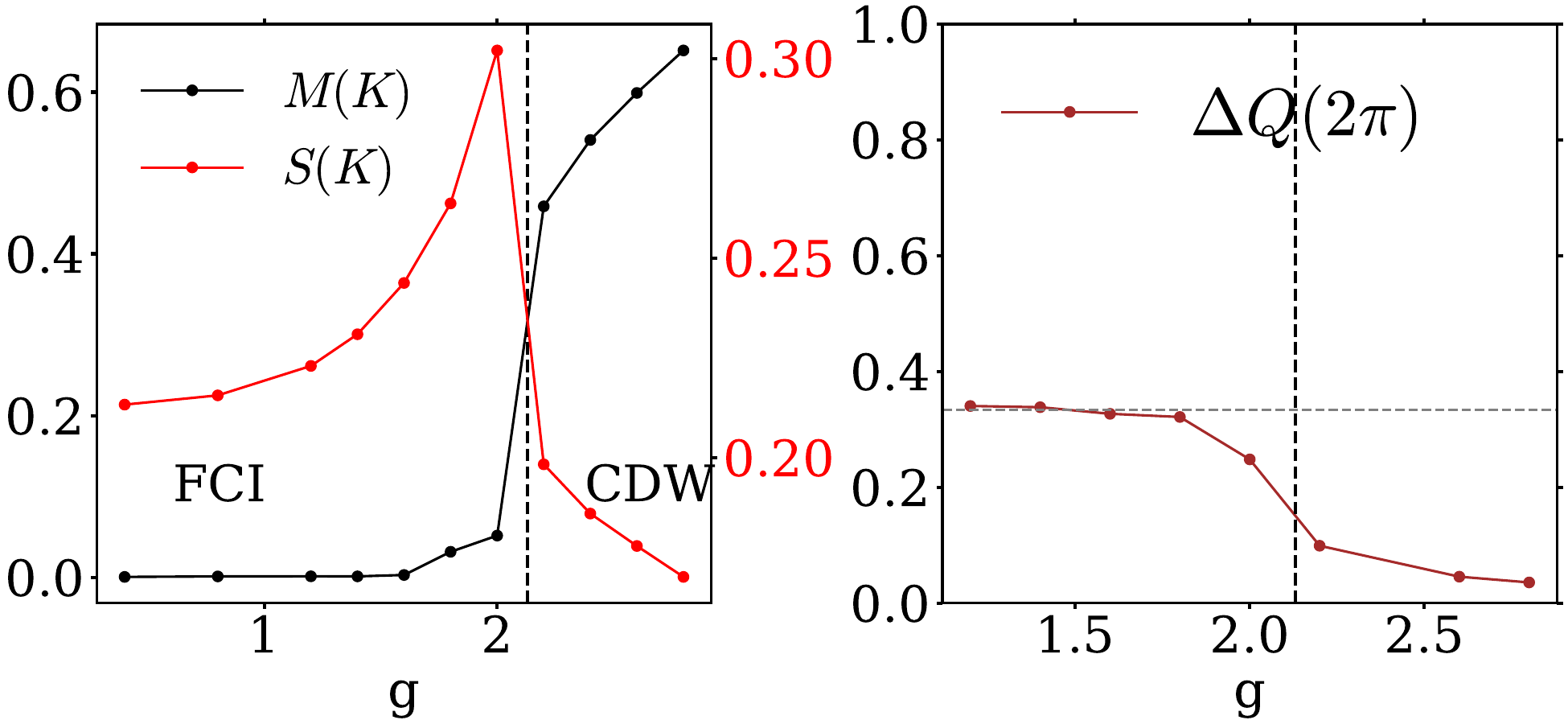}
    \put(-5,46){(a)}
    \put(48,46){(b)}
    \end{overpic}
    \begin{overpic}
        [width=\linewidth]{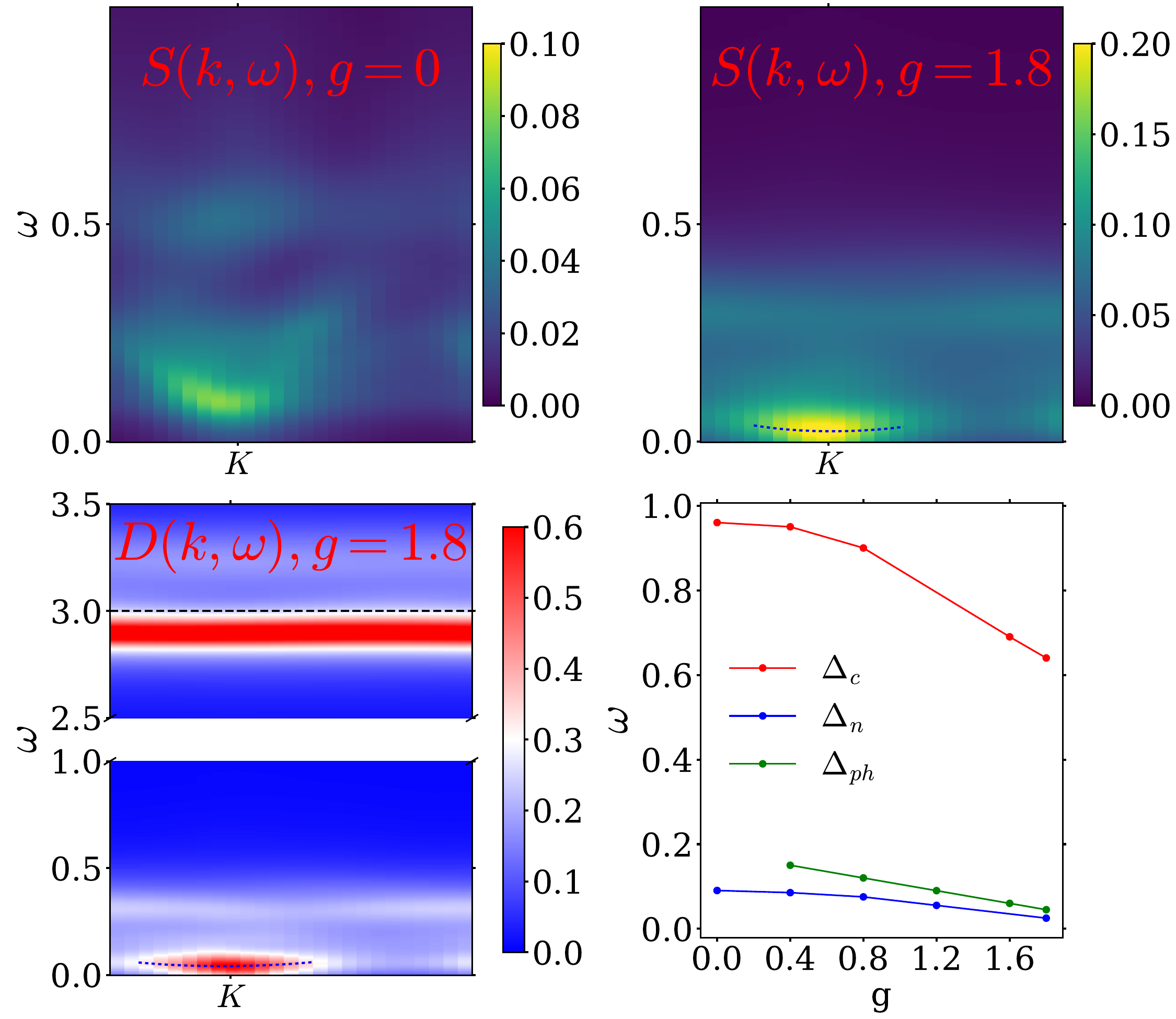}
    \put(0,85){(c)}
    \put(49,85){(d)}
    \put(0,44){(e)}
    \put(49,44){(f)}
    \end{overpic}
    
    \caption{\textbf{FCI-CDW transition and Magnetoroton-polarons}. The spectrum shown for the FCI case focuses along path 2 in Fig.~\ref{fig:fig1} (d). (a) evolution of CDW parameter $M(K)$ and static density correlation $S(K)$ at $\nu = 1/3$ filling as $g$ increases. (b) Charge pumping by flux threading. (c) and (d) show $S(k,\omega)$ defined in Eq.~\eqref{eq:spectra function} for $g = 0$ and $g = 1.8$ close to the phase boundary; a dashed line is added to guide the magnetoroton-polaron dispersion in (d) and (e). (e) Phonon spectral function $D(k,\omega)$; the black dashed line denotes the intrinsic frequency $\omega_0 = 3$. We use a broken $\omega$ axis here. (f) Evolution of neutral gap $\Delta_n$, charge gap $\Delta_c$, and phonon gap $\Delta_{ph}$ within the FCI phase.}
    \label{fig:fig3}
\end{figure}

In this work, we address the above question by studying a prototypical model, the flat-band Haldane model coupled to dynamical Holstein phonons, filled with spinless fermions at integer and fractional fillings via state-of-the-art density matrix renormalization group (DMRG) with local basis optimization(LBO) ~\cite{Zhang1998Density, Brockt2015Matric}. As schematically shown in Fig.~\ref{fig:fig1}, we demonstrate that for both CI and FCI, the low-lying density excitations are strongly dressed by phonons, forming collective particles dubbed exciton-polarons and magnetoroton-polarons. In contrast to the formation of graviton polaritons in a cavity setting~\cite{bacciconi_prx2025_gravitonpolaritons}, we find that electron–phonon coupling amplifies the chirality deficit of the graviton response. 
%making it more difficult to detect in lattice FCI than in its Landau-level(LL) counterparts. 
Interestingly, the phonons imprint the dispersion of those excitations in their spectrum, as shown in Fig.~\ref{fig:fig2}(e) and Fig.~\ref{fig:fig3} (e). When the coupling strength is strong, the charge order emerges accompanied by the breakdown of Hall conductivity quantization. Across the transitions, the charge gap remains intact and larger than the charge-neutral gap, suggesting a unified neutral-mode-instability-triggered transition in both the CI and FCI cases.

\newsect{Model and Method}
We study spinless fermions on the honeycomb lattice described by the extended Haldane-Holstein model[Fig.~\ref{fig:fig1}(c)]~\cite{haldane1988model}
\begin{equation}
\begin{aligned}
H=&-t_1\sum_{\langle ij\rangle}(c_i^\dagger c_j+\mathrm{H.c.})
-t_2\sum_{\langle\langle ij\rangle\rangle}
(e^{i\phi_{ij}}c_i^\dagger c_j+\mathrm{H.c.}) \\
&-t_3\sum_{\langle\langle\langle ij\rangle\rangle\rangle}
(c_i^\dagger c_j+\mathrm{H.c.})
+V_1\sum_{\langle ij\rangle}n_i n_j
\\
&+\omega_0\sum_i b_i^\dagger b_i  + g\omega_0\sum_i(b_i+b_i^\dagger)(n_i-\bar{n})
\end{aligned}
\label{eq:model}
\end{equation}
where $t_1=1$ sets the energy scale, $\bar{n} = \nu/2$ is the averaged charge density per site with $\nu$ the filling fraction of lower band, and we choose $t_2=0.6$, $t_3=-0.54$, $\phi=0.4\pi$, and $V_1=0.5$ for $\nu = 1$ filling and $V_1 = 4$ for $\nu = 1/3$ filling, yielding a nearly flat Chern band with band gap $\Delta_b \sim 2.5$ and banwidth $W \sim 0.25$~\cite{Neupert2011_flat_chern_band}.

To obtain the spectrum, we first compute ground states using DMRG~\cite{White1992,White1993,Schollwock2011} with LBO~\cite{Zhang1998Density, Brockt2015Matric} (which gives a Tree Tensor Network to encode the ground state; see Supplemental Materials (SM)~\cite{suppmat} for details). We focus on $L_y=3$ cylinders, retaining up to $D=1200$ states and truncating the local phonon occupation at $n_{\rm ph}^{\max}=4$. Dynamical response functions are calculated using TDVP~\cite{Haegeman2011,Haegeman2016,Bauernfeind2020Time}, primarily on $L_x=16$ cylinders, with selected calculations performed at $L_x=24$ to reduce finite-size effects.  Energy spectra are obtained by Fourier transforming the corresponding real-space and real-time correlation functions. SM~\cite{suppmat} provides the convergence check. 

The spectral function for density is computed as:
\begin{equation}
    \begin{aligned}
        S(\mathbf{k},\omega) &= \frac{1}{\sqrt{N_tN_A}}  \sum_{jl} \ e^{i(\omega+i\eta) t_l } e^{-i\mathbf{k}(\mathbf{r_j-r_0})} \\ &\big( \langle n_{j,A}(t_l)n_{0,A}\rangle 
        -\langle n_{j,A}\rangle \langle n_{0,A}\rangle\big).
    \end{aligned}
    \label{eq:spectra function}
\end{equation}
Here we only count the correlation in the A sublattice, as the FCI phase is sublattice symmetric and we set $\eta = 0.05$.  The phonon spectral function, denoted by \(D(\mathbf{k},\omega)\), is defined analogously. The momentum path is quantized due to the finite $L_y$ in the cylinder simulation, as shown in Fig.~\ref{fig:fig1}(d).

We also apply twisted boundary conditions to measure the response of the ground state to external flux; this gives the Hall conductivity within the FCI phase [Fig.~\ref{fig:fig3}(b)].

\newsect{Exciton-polaron in CI phase}
At integer filling, electron–phonon coupling has primarily been studied as a mechanism for destroying the band topology of Chern insulators in the strong-coupling regime~\cite{Cangemi2019Topological,SousaJnior2026Realspace}. When the coupling strength is mild, the CI phase is robust and protected by the topological gap, whose energy scale is set by the band gap. The strong-coupling limit can be solved using the Lang-Firsov transformation~\cite{mahan1981many} (see SM~\cite{suppmat} for the details), where the electrons are fully localized. Therefore, there will be a phase transition between the CI and the fully localized phase, which has now been identified as a first-order phase transition~\cite{SousaJnior2026Realspace}.

Here we study the excitations within the CI phase. The low-lying charge-neutral excitation in CI involves inter-band dynamics. It is a bound excitation of an electron and a hole (as depicted in Fig.~\ref{fig:fig1} (a)), which has its minimum located at the $\Gamma$ point~\cite{linExciton2022,Hongyu_prl2024_thermodinamicFCI}. Before we investigate the excitations, we first check the endurance of the topological phase against the electron-phonon coupling. Fig.~\ref{fig:fig2} (a) shows the evolution of charge order and correlation at the $\Gamma$ point, evaluated using 
\begin{equation}
\begin{aligned}
     &M(\mathbf{k})= \sum_i e^{-i\mathbf{k}\cdot \mathbf{r_i}} \langle n^A_i - \bar{n} \rangle / N, \\ &
     S(\mathbf{k}) = \sum_{ij} e^{-i\mathbf{k} \cdot (\mathbf{r_i}-\mathbf{r_j)}}(\langle n_i^A n_j^A\rangle - \langle n_i^A \rangle\langle n_j^A \rangle) / N.
\end{aligned}
\end{equation}
As the coupling strength increases, the density fluctuation is enhanced, followed by the emergence of charge order at $\Gamma$, suggesting that the phase instability is related to density fluctuations. We mark $g \sim 1.6$ as the critical value.

Fig.~\ref{fig:fig2} (b) and (c) show the spectral function $S(\mathbf{k},\omega)$ at different values of $g$, where the exciton appears as a sharp mode at the $\Gamma$ point. As $g$ increases, this mode gains spectral weight and softens, demonstrating phonon dressing and the formation of an exciton-polaron. Moreover, the exciton-polaron dispersion becomes visible in the phonon spectrum [Fig.~\ref{fig:fig2} (e)], while there is residual weight around the phonon intrinsic frequency $\omega_0 = 3$, the most spectrum weight is transferred to the exciton-polaron dispersion line and concentrates at its minimum, signaling strong reconstruction of the phonon mode.  

The effect of electron-phonon coupling can be understood as a Semenoff mass term in the language of Dirac fermions ~\cite{Semenoff1984Condensed}. Starting with electron-phonon coupling term $H_{ep} = g\omega_0 \sum_{r,\alpha}(b_{r,\alpha}+b_{r,\alpha}^\dagger)(n_{r,\alpha}-\bar{n})$, where the site index is decomposed into unit cell indices and sublattice indices ($\alpha \in \{A,B\}$),  denote $X_{r,\alpha} = b_{r,\alpha}+b_{r,\alpha}^\dagger$ as the phonon displacement operator and $\delta n_{r,\alpha} =n_{r,\alpha} - \bar{n}$, the coupling term could be rewritten as
\begin{equation}
\end{equation}
where $X_r^\pm = X_{r,A} \pm X_{r,B}$ and similar for $\delta n^\pm_r$. Then $\langle \delta n_r^+ \rangle = 0$, and in the first-order approximation only $X_r^-  \delta n^-_r$ remains, where $X_r^-$ acts as a dynamical mass term but with an alternating sign between $A,B$ sublattice, therefore favoring sublattice polarization.  We further resolve the spectral function for $n^-$ branches by replacing the observable $n_{r,A}$ with $n_{r,-}$ in Eq.~\eqref{eq:spectra function}. Indeed, the low energy branch in Fig.~\ref{fig:fig2} (c) is fully captured by $S^-(\mathbf{k},\omega)$ shown in Fig.~\ref{fig:fig2} (d).

We summarize the evolution of charge $\Delta_c$, density $\Delta_n$ and phonon $\Delta_{ph}$ gaps in Fig.~\ref{fig:fig2} (f), $\Delta_{ph}$ follows closely with $\Delta_n$ and exhibiting a polaron behavior. While the charge gap is also renormalized by phonons, it remains larger than the density gap and less sensitive to phonon coupling. We also explore the role of phonon frequency and find that changing the phonon frequency renormalizes the exciton-polaron energy, and we do not observe avoid splitting behavior between phonons and excitons (see SM~\cite{suppmat} for the result).

\newsect{Magnetoroton-polaron in FCI}
We now turn to the fractionally filled case. The low-energy density excitation of the FCI is the magnetoroton, a neutral bound state of a fractionally charged quasiparticle and quasihole, whose minimum occurs at finite momentum. We saw that phonons strongly dress the exciton to form the exciton-polaron in the CI. Here, we show that phonons also dress the magnetoroton and form magnetoroton-polarons, as depicted in Fig.~\ref{fig:fig1} (b).

As in the CI case, we first examine the stability of the FCI ground state against electron-phonon coupling. For small $g$, the CDW order parameter remains negligible [Fig.~\ref{fig:fig3} (a)], while $2\pi$ flux threading pumps charge $\Delta Q(2\pi)$, which gives a fractionally quantized Hall conductivity within the FCI phase $\sigma_{xy}=1/3$   [Fig.~\ref{fig:fig3} (b)], confirming that the ground state remains in the FCI phase. As $g$ approaches the critical value $g_c\simeq 2.1$, the static density correlation $S(K)$ is strongly enhanced, followed by the rapid development of charge order $M(K)$. Concurrently, $\Delta Q(2\pi)$ departs from the quantized value of $1/3$[Fig.~\ref{fig:fig3} (b)] . These results establish a transition from the FCI to a CDW phase with ordering wave vector $K$.

Having established the FCI regime, we next examine its excitation spectrum. In the absence of electron--phonon coupling, $S(\mathbf{k},\omega)$ exhibits a dispersive magnetoroton branch with a pronounced minimum at the $K$ point [Fig.~\ref{fig:fig3} (c)]. Upon increasing $g$ toward the phase boundary, this mode gains spectral weight and softens substantially [Fig.~\ref{fig:fig3} (d)]. The enhancement of $S(K)$ near the transition is therefore closely connected to the softening of the magnetoroton minimum.

Similar to the CI case, the phonon spectrum provides direct evidence for strong dressing of the neutral mode. In addition to the residual spectral weight near the bare phonon frequency $\omega_0$, $D(\mathbf{k},\omega)$ develops a low-energy dispersive branch that closely follows the magnetoroton dispersion [Fig.~\ref{fig:fig3} (e)]. In particular, the phonon spectral weight is concentrated near the magnetoroton minimum at $K$. The matching dispersions in the electronic and phononic responses demonstrate that phonons dress the magnetoroton into a magnetoroton-polaron. Unlike a conventional polaron formed from a single charged carrier, this composite excitation originates from a neutral bound state of fractionally charged quasiparticles.

The evolution of the excitation gaps further supports this interpretation. As shown in Fig.~\ref{fig:fig3}(f), the phonon gap $\Delta_{\mathrm{ph}}$ closely follows the neutral gap $\Delta_n$, and both are strongly suppressed as $g$ increases. By contrast, although the charge gap $\Delta_c$ is also renormalized, it remains much larger and finite throughout the FCI regime. This separation of energy scales indicates that phonons predominantly dress the neutral density sector. Together with the enhancement of $S(K)$ and the emergence of CDW order at the same wave vector, these results support a magnetoroton-polaron-trigger instability of the FCI. Remarkably, this phonon-driven crystallization occurs at dilute filling with only nearest-neighbor electron--electron interactions and onsite Holstein coupling. Although the FCI remains stable at $g = 0$ even when the interaction strength substantially exceeds the single-particle band gap~\cite{Kourtis2014Fractional}, coupling to dynamical phonons ultimately drives it into a CDW phase. 

Same as CI case, we also check the effect of phonon frequency and find that changing the phonon frequency renormalizes the magnetoroton-polaron energy, and we do not observe avoid splitting behavior between phonons and excitons (see SM~\cite{suppmat} for the result).

\newsect{chirality deficit of graviton mode in FCI}
\begin{figure}
    \centering
    \begin{overpic}
        [width=\linewidth]{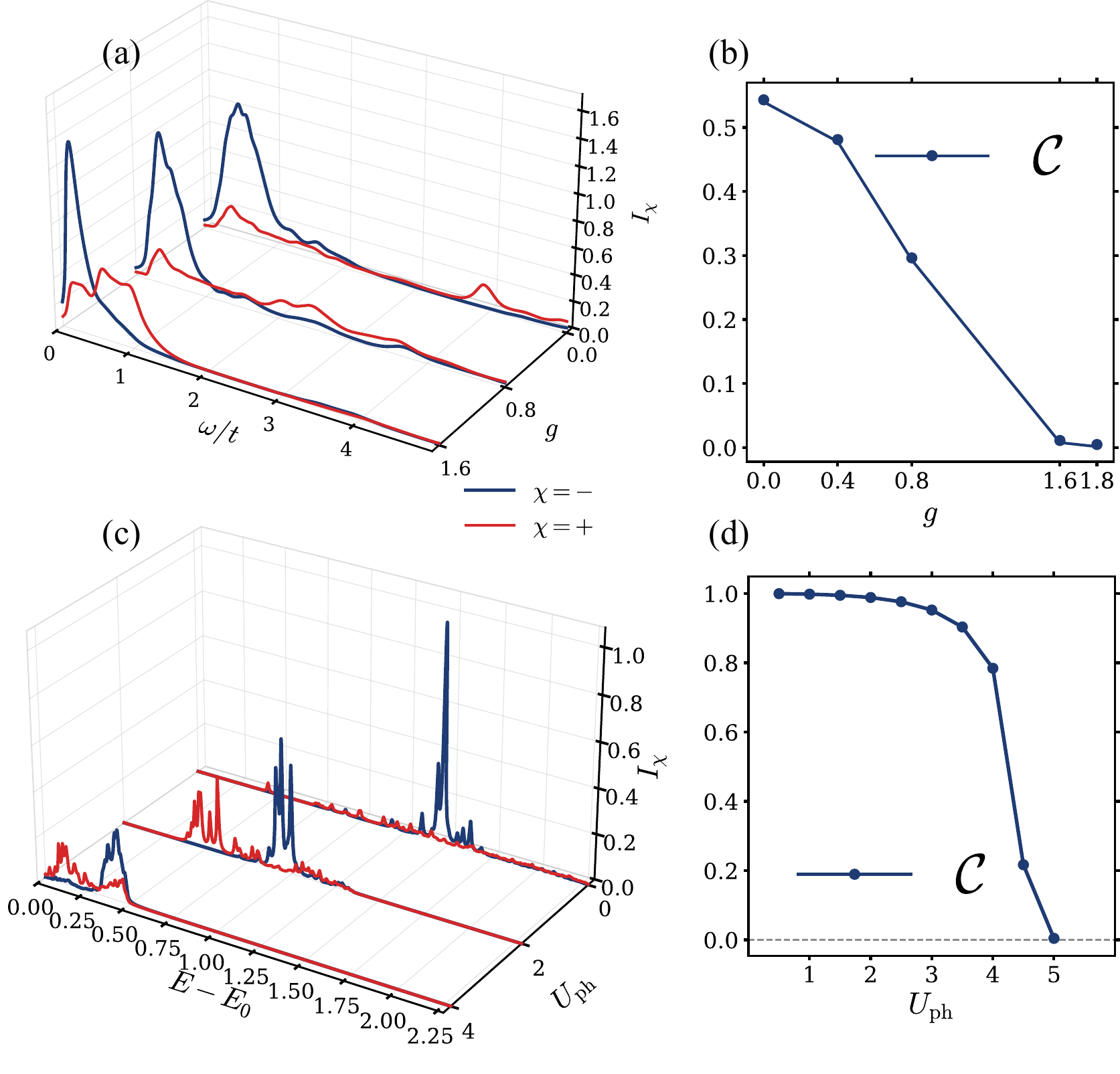}
    \end{overpic}
    \caption{
\textbf{Phonon-induced chirality deficit of the graviton response.}
(a) Chiral graviton spectra $I_{\chi}(\omega)$ in the FCI for
electron--phonon couplings $g=0$, $0.8$, and $1.6$, calculated on an
$L_x\times L_y=16\times3$ cylinder. The blue solid and red dashed
curves denote the $\chi=-$ and $\chi=+$ channels, respectively; the
spectra are normalized by the total weight in $\chi = -$ sector. (b)Chirality ratio $\mathcal{C} = (N_--N_{+})/(N_{-}+N_+)$ of the integrated
spectral weights in the two chiral channels as a function of $g$.
(c) Chiral graviton spectra obtained from a complementary
Landau-level calculation with a phonon-mediated effective interaction
of strength $U_{\mathrm{ph}}$. The spectra are plotted against $E-E_0$, where $E_0$ is the ground state energy. (d) Corresponding ratio $\mathcal{C}$ as a function of
$U_{\mathrm{ph}}$.
}
    \label{fig:fig4}
\end{figure}
The chiral graviton mode (CGM) is a genuine feature of FQH states~\cite{haldane2011geometrical}; it is a long-wavelength excitation and describes the chiral deformation of the correlation hole and can be measured through the Landau Level metric tensor~\cite{nguyen2014lowest,liou2019chiral}.

Interestingly, acoustic waves were proposed as a probe of the quadrupolar collective mode in fractional quantum Hall liquids~\cite{yang2016acoustic}. Here we investigate how dynamical phonons affect the CGM by examining its response to a chiral quadrupolar density operator~\cite{longSpectra2025,xavier2025chiralgravitonslattice,longChiral2026,bacciconiChiral2026},
\begin{equation}
    \begin{aligned}
    O^{\pm} & =\sum_{r_i,\delta} e^{\pm 2i \arg{[\delta]}}f(\delta) n_{r_i}n_{r_i+\delta}, 
    \end{aligned}
\label{eq:definition_Onn}
\end{equation}
where $\arg{[\delta]}$ is the angle given by the vector $\delta = \vec{r}_i-\vec{r}_j$ and $f_G(\delta)$ a short ranged function (here $f(\delta)= 1/|\delta|$ for up to NNN and 0 otherwise). We measure the spectrum function $I^\pm(\omega)$ similarly to Eq.~\eqref{eq:spectra function} but without a spatial Fourier transform, since Eq.~\eqref{eq:definition_Onn} already defines an excitation at the $\Gamma$ point.

Phonons can modify graviton chirality by renormalizing the electronic interaction. Fig.~\ref{fig:fig4} (a) shows the evolution of the circular quadrupolar response with electron-phonon coupling. At $g=0$, the spectrum is dominated by a pronounced negative-helicity resonance, while the positive-helicity response is much weaker. As $g$ increases, the opposite-helicity contribution becomes more prominent relative to the negative-helicity response when approaching the phase boundary [Fig~\ref{fig:fig3}(b),Fig~\ref{fig:fig4}(b)] (We quantify this effect using the chirality ratio
$\mathcal{C}=(N_--N_+)/(N_-+N_+)$, as shown in
Fig.~\ref{fig:fig4}(b)), demonstrating a strong chirality deficit effect. To understand the origin of this change, we consider a continuum FQH system in the lowest Landau level at $\nu=1/3$. In this case, virtual phonon exchange generates a correction to the electron-electron interaction (see SM~\cite{suppmat}). This correction deforms the ground-state correlations away from those of the model Laughlin state. Fig.~\ref{fig:fig4} (c) shows that the circular-response chirality can be strongly suppressed by the phonon-mediated interaction. The representative spectrum in Fig.~\ref{fig:fig4} (c) exhibits a nearly complete loss of the graviton chirality, even though the ground state remains in the FQH regime. Thus, the graviton chirality is sensitive to the microscopic ground-state correlations perturbed by phonon couplings even when the system remains in the same FQH phase, whereas the cavity mode strongly dresses the graviton mode into graviton polaritons~\cite{bacciconi_prx2025_gravitonpolaritons}.

\newsect{Discussion} 
In summary, using DMRG with LBO, we have shown that Holstein phonons predominantly dress the low-energy charge-neutral excitations of integer and fractional Chern insulators. These results establish a unified physical picture of phonon dressing in CIs and FCIs and demonstrate that phonons both reveal neutral collective modes in the lattice response and control their softening toward charge-ordering instabilities.  In both CI and FCI, the phonon spectrum inherits the dispersion of the corresponding electronic collective mode, providing a direct signature of polaron formation. 
We also find a chirality deficit within the FCI phase, where the phonon enhances the spectrum weight from the opposite-chirality channel.

Beyond charge order, phonon-mediated interactions may favor competing
pairing tendencies. Recent observations
of superconductivity proximate to fractional topological and anomalous
Hall regimes in twisted TMD moiré systems and rhombohedral multilayer
graphene further motivate investigating the role of phonons in
superconductivity and its interplay with fractional topology and charge
order in spinful or multiband Chern systems~\cite{Han2025SignatureChiralSC,Xia2024SuperconductivityWSe2,sun2026twistangleevolutionvalleypolarizedfractional,Choi2025SuperconductivityQAHRhombohedra,Kumar2026SuperconductivityRhombohedra,lu2024fractionalQHmultilayergraphene}.  
 Our results may also have finite-temperature implications: phonon-induced magnetoroton softening provides a microscopic route to the enhanced bulk entropy proposed to drive thermal transitions between competing quantum Hall regimes~\cite{Kim2026entropy}. Our results further suggest that lattice vibrations may complicate the identification of the graviton response in lattice systems~\cite{wangDynamics2025,longChiral2026,shenMagnetorotons2024}.

Several questions remain open. An immediate direction is to determine the nature of the phonon-driven FCI--CDW transition, whether it is first order or continuous. The difference in the fate of the graviton mode between cavity coupling and electron-phonon coupling is quite intriguing. We hope a single-mode treatment of magnetoroton-- phonon coupling could help clarify this. It will also be important to extend the present study beyond local Holstein phonons to dispersive phonon branches and nonlocal Fröhlich-type electron--phonon coupling, where momentum-dependent coupling may selectively dress different collective modes. Finally, extending the present analysis to material-specific models of twisted transition metal dichalcogenide and rhombohedral pentalayer graphene represents a promising direction.

\paragraph{\it Acknowledgement} We acknowledge Hongyu Lu for carefully reading the manuscript and for helpful discussions throughout the course of this project. ML and ZYM thank Eric Jeckelmann, João C Inácio, Natanael C Costa, Kai Sun, Fakher Assaad for the helpful discussions, and acknowledge  the support from the Research Grants Council (RGC) of Hong Kong (Project Nos.
C7037-22GF, 17302223, 17301924, 17301725, 17302026, AoE/P-604/25-R), the State Key Laboratory of Optical Quantum Materials
at HKU, and the ANR/RGC Joint Research Scheme sponsored by RGC of Hong Kong (Project No. A\_HKU703/22).
We thank the HPC2021 system under the Information Technology Services at the University of Hong Kong~\cite{hpc2021}, as well as Beijing Paratera Tech Corp., Ltd~\cite{paratera} for providing HPC resources that have contributed to the research results reported within this paper.  The code developed in this paper is based on the TensorKit package~\cite{Tensorkit_web}.

\bibliographystyle{longapsrev4-2}
\bibliography{bib}

@book{mahan1981many,
  title={Many-particle physics},
  author={Mahan, Gerald D},
  year={1981},
  publisher={Plenum Press},
  address={New York}
}

@misc{Kim2026entropy,
      title={Entropy-driven transitions between extended integer and fractional quantum Hall regimes}, 
      author={Kyung-Su Kim and Steven A. Kivelson},
      year={2026},
      eprint={2609.16483},
      archivePrefix={arXiv},
      primaryClass={cond-mat.mes-hall},
      url={https://arxiv.org/abs/2609.16483}, 
}

@article{Kourtis2014Fractional,
  title = {Fractional Chern Insulators with Strong Interactions that Far Exceed Band Gaps},
  volume = {112},
  ISSN = {1079-7114},
  url = {http://dx.doi.org/10.1103/PhysRevLett.112.126806},
  DOI = {10.1103/physrevlett.112.126806},
  number = {12},
  journal = {Physical Review Letters},
  publisher = {American Physical Society (APS)},
  author = {Kourtis,  Stefanos and Neupert,  Titus and Chamon,  Claudio and Mudry,  Christopher},
  year = {2014},
  month = Mar 
}

@article{Xiao2011_quantum_hall,
	title = {Interface engineering of quantum Hall effects in digital transition metal oxide heterostructures},
	author = {Xiao, Di
and Zhu, Wenguang
and Ran, Ying
and Nagaosa, Naoto
and Okamoto, Satoshi},
	journal = {Nature Communications},
	volume = {2},
	issue = {1},
	pages = {596},
	year = {2011},
	month = {Dec},
	doi = {10.1038/ncomms1602},
	url = {https://doi.org/10.1038/ncomms1602}
}

@article{Bo2020Microscopic,
  title = {Microscopic theory for nematic fractional quantum Hall effect},
  author = {Yang, Bo},
  journal = {Phys. Rev. Res.},
  volume = {2},
  issue = {3},
  pages = {033362},
  numpages = {11},
  year = {2020},
  month = {Sep},
  publisher = {American Physical Society},
  doi = {10.1103/PhysRevResearch.2.033362},
  url = {https://link.aps.org/doi/10.1103/PhysRevResearch.2.033362}
}

@article{Xia2024SuperconductivityWSe2,
  title = {Superconductivity in twisted bilayer WSe2},
  volume = {637},
  ISSN = {1476-4687},
  url = {http://dx.doi.org/10.1038/s41586-024-08116-2},
  DOI = {10.1038/s41586-024-08116-2},
  number = {8047},
  journal = {Nature},
  publisher = {Springer Science and Business Media LLC},
  author = {Xia,  Yiyu and Han,  Zhongdong and Watanabe,  Kenji and Taniguchi,  Takashi and Shan,  Jie and Mak,  Kin Fai},
  year = {2024},
  month = Oct,
  pages = {833–838}
}

@article{Choi2025SuperconductivityQAHRhombohedra,
  title = {Superconductivity and quantized anomalous Hall effect in rhombohedral graphene},
  volume = {639},
  ISSN = {1476-4687},
  url = {http://dx.doi.org/10.1038/s41586-025-08621-y},
  DOI = {10.1038/s41586-025-08621-y},
  number = {8054},
  journal = {Nature},
  publisher = {Springer Science and Business Media LLC},
  author = {Choi,  Youngjoon and Choi,  Ysun and Valentini,  Marco and Patterson,  Caitlin L. and Holleis,  Ludwig F. W. and Sheekey,  Owen I. and Stoyanov,  Hari and Cheng,  Xiang and Taniguchi,  Takashi and Watanabe,  Kenji and Young,  Andrea F.},
  year = {2025},
  month = Mar,
  pages = {342–347}
}

@article{lu2024fractionalQHmultilayergraphene,
  title={Fractional quantum anomalous Hall effect in multilayer graphene},
  author={Lu, Zhengguang and Han, Tonghang and Yao, Yuxuan and Reddy, Aidan P and Yang, Jixiang and Seo, Junseok and Watanabe, Kenji and Taniguchi, Takashi and Fu, Liang and Ju, Long},
  journal={Nature},
  volume={626},
  number={8000},
  pages={759--764},
  year={2024},
  publisher={Nature Publishing Group UK London}
}

@article{Kumar2026SuperconductivityRhombohedra,
  title = {Superconductivity from dual-surface carriers in rhombohedral graphene},
  volume = {22},
  ISSN = {1745-2481},
  url = {http://dx.doi.org/10.1038/s41567-026-03277-5},
  DOI = {10.1038/s41567-026-03277-5},
  number = {6},
  journal = {Nature Physics},
  publisher = {Springer Science and Business Media LLC},
  author = {Kumar,  Manish and Waleffe,  Derek and Okounkova,  Anna and Tejani,  Raveel and Phong,  Võ Tiến and Watanabe,  Kenji and Taniguchi,  Takashi and Lewandowski,  Cyprian and Folk,  Joshua and Yankowitz,  Matthew},
  year = {2026},
  month = June,
  pages = {862–869}
}

@article{Han2025SignatureChiralSC,
  title = {Signatures of chiral superconductivity in rhombohedral graphene},
  volume = {643},
  ISSN = {1476-4687},
  url = {http://dx.doi.org/10.1038/s41586-025-09169-7},
  DOI = {10.1038/s41586-025-09169-7},
  number = {8072},
  journal = {Nature},
  publisher = {Springer Science and Business Media LLC},
  author = {Han,  Tonghang and Lu,  Zhengguang and Hadjri,  Zach and Shi,  Lihan and Wu,  Zhenghan and Xu,  Wei and Yao,  Yuxuan and Cotten,  Armel A. and Sharifi Sedeh,  Omid and Weldeyesus,  Henok and Yang,  Jixiang and Seo,  Junseok and Ye,  Shenyong and Zhou,  Muyang and Liu,  Haoyang and Shi,  Gang and Hua,  Zhenqi and Watanabe,  Kenji and Taniguchi,  Takashi and Xiong,  Peng and Zumb\"{u}hl,  Dominik M. and Fu,  Liang and Ju,  Long},
  year = {2025},
  month = May,
  pages = {654–661}
}

@misc{sun2026twistangleevolutionvalleypolarizedfractional,
      title={Twist-angle evolution from valley-polarized fractional topological phases to valley-degenerate superconductivity in twisted bilayer MoTe2}, 
      author={Zheng Sun and Fan Xu and Jiayi Li and Yifan Jiang and Jingjing Gao and Cheng Xu and Tongtong Jia and Kehao Cheng and Jinyang Zhang and Wanghao Tian and Kenji Watanabe and Takashi Taniguchi and Jinfeng Jia and Shengwei Jiang and Yang Zhang and Yuanbo Zhang and Shiming Lei and Xiaoxue Liu and Tingxin Li},
      year={2026},
      eprint={2603.16412},
      archivePrefix={arXiv},
      primaryClass={cond-mat.mes-hall},
      url={https://arxiv.org/abs/2603.16412}, 
}

@misc{suppmat,
  note = {See Supplemental Material for the Lang-Firsov transformation in the strong-coupling limit, Tree tensor network optimization and time evolution, endurance of Hall conductivity against electron-phonon coupling, convergence check, spectrum on all allowed momentum paths, finite-size extrapolation, effect of phonon frequency, and details on the perturbative treatment of electron-phonon interaction in Landau Level.}
}

@misc{Tensorkit_web,
    title = {TensorKit},
    howpublished = {\url{https://jutho.github.io/TensorKit.jl/stable/}},
    note = {Accessed: December 21, 2024}
}

@article{yang2016acoustic,
  title = {{Acoustic wave absorption as a probe of dynamical geometrical response of fractional quantum Hall liquids}},
  author = {Yang, Kun},
  journal = {Phys. Rev. B},
  volume = {93},
  issue = {16},
  pages = {161302},
  numpages = {4},
  year = {2016},
  month = {Apr},
  publisher = {American Physical Society},
  doi = {10.1103/PhysRevB.93.161302},
  url = {https://link.aps.org/doi/10.1103/PhysRevB.93.161302}
}

@article{pu2024microscopic,
  title = {{Microscopic Model for Fractional Quantum Hall Nematics}},
  author = {Pu, Songyang and Balram, Ajit C. and Taylor, Joseph and Fradkin, Eduardo and Papi\ifmmode \acute{c}\else \'{c}\fi{}, Zlatko},
  journal = {Phys. Rev. Lett.},
  volume = {132},
  issue = {23},
  pages = {236503},
  numpages = {9},
  year = {2024},
  month = {Jun},
  publisher = {American Physical Society},
  doi = {10.1103/PhysRevLett.132.236503},
  url = {https://link.aps.org/doi/10.1103/PhysRevLett.132.236503}
}

@article{laughlin1983anomalous,
  title = {{Anomalous Quantum Hall Effect: An Incompressible Quantum Fluid with Fractionally Charged Excitations}},
  author = {Laughlin, R. B.},
  journal = {Phys. Rev. Lett.},
  volume = {50},
  issue = {18},
  pages = {1395--1398},
  numpages = {0},
  year = {1983},
  month = {May},
  publisher = {American Physical Society},
  doi = {10.1103/PhysRevLett.50.1395},
  url = {https://link.aps.org/doi/10.1103/PhysRevLett.50.1395}
}

@article{gmp1986magnetoroton,
  title = {{Magneto-roton theory of collective excitations in the fractional quantum Hall effect}},
  author = {Girvin, S. M. and MacDonald, A. H. and Platzman, P. M.},
  journal = {Phys. Rev. B},
  volume = {33},
  issue = {4},
  pages = {2481--2494},
  numpages = {0},
  year = {1986},
  month = {Feb},
  publisher = {American Physical Society},
  doi = {10.1103/PhysRevB.33.2481},
  url = {https://link.aps.org/doi/10.1103/PhysRevB.33.2481}
}

@article{haldane2011geometrical,
  title = {{Geometrical Description of the Fractional Quantum Hall Effect}},
  author = {Haldane, F. D. M.},
  journal = {Phys. Rev. Lett.},
  volume = {107},
  issue = {11},
  pages = {116801},
  numpages = {5},
  year = {2011},
  month = {Sep},
  publisher = {American Physical Society},
  doi = {10.1103/PhysRevLett.107.116801},
  url = {https://link.aps.org/doi/10.1103/PhysRevLett.107.116801}
}

@article{You2014TheoryNematic,
  title = {Theory of Nematic Fractional Quantum Hall States},
  volume = {4},
  ISSN = {2160-3308},
  url = {http://dx.doi.org/10.1103/PhysRevX.4.041050},
  DOI = {10.1103/physrevx.4.041050},
  number = {4},
  journal = {Physical Review X},
  publisher = {American Physical Society (APS)},
  author = {You,  Yizhi and Cho,  Gil Young and Fradkin,  Eduardo},
  year = {2014},
  month = Dec 
}

@article{haldane1988model,
  title={Model for a quantum Hall effect without Landau levels: Condensed-matter realization of the" parity anomaly"},
  author={Haldane, F Duncan M},
  journal={Physical review letters},
  volume={61},
  number={18},
  pages={2015},
  year={1988},
  publisher={APS}
}

@article{Semenoff1984Condensed,
  title = {Condensed-Matter Simulation of a Three-Dimensional Anomaly},
  volume = {53},
  ISSN = {0031-9007},
  url = {http://dx.doi.org/10.1103/PhysRevLett.53.2449},
  DOI = {10.1103/physrevlett.53.2449},
  number = {26},
  journal = {Physical Review Letters},
  publisher = {American Physical Society (APS)},
  author = {Semenoff,  Gordon W.},
  year = {1984},
  month = Dec,
  pages = {2449–2452}
}

@article{Appugliese2022Breakdown,
  title = {Breakdown of topological protection by cavity vacuum fields in the integer quantum Hall effect},
  volume = {375},
  ISSN = {1095-9203},
  url = {http://dx.doi.org/10.1126/science.abl5818},
  DOI = {10.1126/science.abl5818},
  number = {6584},
  journal = {Science},
  publisher = {American Association for the Advancement of Science (AAAS)},
  author = {Appugliese,  Felice and Enkner,  Josefine and Paravicini-Bagliani,  Gian Lorenzo and Beck,  Mattias and Reichl,  Christian and Wegscheider,  Werner and Scalari,  Giacomo and Ciuti,  Cristiano and Faist,  Jér\^ome},
  year = {2022},
  month = Mar,
  pages = {1030–1034}
}

@article{nguyen2014lowest,
	author = {Nguyen, Dung Xuan and Son, Dam Thanh and Wu, Chaolun},
	title = {{Lowest Landau Level Stress Tensor and Structure Factor of Trial Quantum Hall Wave Functions}},
	journal = {arXiv},
	year = {2014},
	month = nov,
	eprint = {1411.3316},
	doi = {10.48550/arXiv.1411.3316}
}

@article{liou2019chiral,
  title = {{Chiral Gravitons in Fractional Quantum Hall Liquids}},
  author = {Liou, Shiuan-Fan and Haldane, F. D. M. and Yang, Kun and Rezayi, E. H.},
  journal = {Phys. Rev. Lett.},
  volume = {123},
  issue = {14},
  pages = {146801},
  numpages = {5},
  year = {2019},
  month = {Sep},
  publisher = {American Physical Society},
  doi = {10.1103/PhysRevLett.123.146801},
  url = {https://link.aps.org/doi/10.1103/PhysRevLett.123.146801}
}

@article{nguyen2022multiple,
  title = {{Multiple Magnetorotons and Spectral Sum Rules in Fractional Quantum Hall Systems}},
  author = {Nguyen, Dung Xuan and Haldane, F. D. M. and Rezayi, E. H. and Son, Dam Thanh and Yang, Kun},
  journal = {Phys. Rev. Lett.},
  volume = {128},
  issue = {24},
  pages = {246402},
  numpages = {6},
  year = {2022},
  month = {Jun},
  publisher = {American Physical Society},
  doi = {10.1103/PhysRevLett.128.246402},
  url = {https://link.aps.org/doi/10.1103/PhysRevLett.128.246402}
}

@article{liang2024evidence,
	author = {Liang, Jiehui and Liu, Ziyu and Yang, Zihao and Huang, Yuelei and Wurstbauer, Ursula and Dean, Cory R. and West, Ken W. and Pfeiffer, Loren N. and Du, Lingjie and Pinczuk, Aron},
	title = {{Evidence for chiral graviton modes in fractional quantum Hall liquids}},
	journal = {Nature},
	volume = {628},
	pages = {78--83},
	year = {2024},
	month = apr,
	issn = {1476-4687},
	publisher = {Nature Publishing Group},
	doi = {10.1038/s41586-024-07201-w}
}

@article{Haegeman2011,
  title = {{Time-Dependent Variational Principle for Quantum Lattices}},
  author = {Haegeman, Jutho and Cirac, J. Ignacio and Osborne, Tobias J. and Pi\ifmmode \check{z}\else \v{z}\fi{}orn, Iztok and Verschelde, Henri and Verstraete, Frank},
  journal = {Phys. Rev. Lett.},
  volume = {107},
  issue = {7},
  pages = {070601},
  numpages = {5},
  year = {2011},
  month = {Aug},
  publisher = {American Physical Society},
  doi = {10.1103/PhysRevLett.107.070601},
  url = {https://link.aps.org/doi/10.1103/PhysRevLett.107.070601}
}

@article{Haegeman2016,
  title = {{Unifying time evolution and optimization with matrix product states}},
  author = {Haegeman, Jutho and Lubich, Christian and Oseledets, Ivan and Vandereycken, Bart and Verstraete, Frank},
  journal = {Phys. Rev. B},
  volume = {94},
  issue = {16},
  pages = {165116},
  numpages = {10},
  year = {2016},
  month = {Oct},
  publisher = {American Physical Society},
  doi = {10.1103/PhysRevB.94.165116},
  url = {https://link.aps.org/doi/10.1103/PhysRevB.94.165116}
}

@article{White1992,
  title = {{Density matrix formulation for quantum renormalization groups}},
  author = {White, Steven R.},
  journal = {Phys. Rev. Lett.},
  volume = {69},
  issue = {19},
  pages = {2863--2866},
  numpages = {0},
  year = {1992},
  month = {Nov},
  publisher = {American Physical Society},
  doi = {10.1103/PhysRevLett.69.2863},
  url = {https://link.aps.org/doi/10.1103/PhysRevLett.69.2863}
}

@article{White1993,
  title = {{Density-matrix algorithms for quantum renormalization groups}},
  author = {White, Steven R.},
  journal = {Phys. Rev. B},
  volume = {48},
  issue = {14},
  pages = {10345--10356},
  numpages = {0},
  year = {1993},
  month = {Oct},
  publisher = {American Physical Society},
  doi = {10.1103/PhysRevB.48.10345},
  url = {https://link.aps.org/doi/10.1103/PhysRevB.48.10345}
}

@article{Schollwock2011,
  title = {{The density-matrix renormalization group in the age of matrix product states}},
  volume = {326},
  ISSN = {0003-4916},
  url = {http://dx.doi.org/10.1016/j.aop.2010.09.012},
  DOI = {10.1016/j.aop.2010.09.012},
  number = {1},
  journal = {Ann. Phys.},
  publisher = {Elsevier BV},
  author = {Schollw\"{o}ck,  Ulrich},
  year = {2011},
  month = jan,
  pages = {96-192}
}

@article{Haegeman2011Time,
  title = {Time-Dependent Variational Principle for Quantum Lattices},
  author = {Haegeman, Jutho and Cirac, J. Ignacio and Osborne, Tobias J. and Pi\ifmmode \check{z}\else \v{z}\fi{}orn, Iztok and Verschelde, Henri and Verstraete, Frank},
  journal = {Phys. Rev. Lett.},
  volume = {107},
  issue = {7},
  pages = {070601},
  numpages = {5},
  year = {2011},
  month = {Aug},
  publisher = {American Physical Society},
  doi = {10.1103/PhysRevLett.107.070601},
  url = {https://link.aps.org/doi/10.1103/PhysRevLett.107.070601}
}

@article{longSpectra2025,
  title = {Spectra of magnetoroton and chiral graviton modes of the fractional Chern insulator},
  volume = {113},
  ISSN = {2469-9969},
  url = {http://dx.doi.org/10.1103/bjrf-b8s9},
  DOI = {Phys. Rev. B 113, L041108},
  number = {4},
  journal = {Physical Review B},
  publisher = {American Physical Society (APS)},
  author = {Long,  Min and Lu,  Hongyu and Wu,  Han-Qing and Meng,  Zi Yang},
  year = {2026},
  month = jan 
}

@ARTICLE{longChiral2026,
doi = {10.1088/1361-6633/ae8369},
url = {https://doi.org/10.1088/1361-6633/ae8369},
year = {2026},
month = {jul},
publisher = {IOP Publishing},
volume = {89},
number = {7},
pages = {078001},
author = {Long, Min and Bacciconi, Zeno and Lu, Hongyu and Xavier, Hernan B and Yang Meng, Zi and Dalmonte, Marcello},
title = {Chiral graviton modes in fermionic fractional Chern insulators},
journal = {Reports on Progress in Physics},
}

@ARTICLE{bacciconiChiral2026,
       author = {{Bacciconi}, Zeno and {Long}, Min and {Xavier}, Hernan and {Lu}, Hongyu and {Dalmonte}, Marcello and {Meng}, Zi Yang},
        title = "{Chiral Graviton Modes in Non-Abelian lattice Fractional Quantum Hall states}",
      journal = {arXiv e-prints},
         year = 2026,
        month = jul,
          eid = {arXiv:2607.06267},
        pages = {arXiv:2607.06267},
          doi = {10.48550/arXiv.2607.06267},
archivePrefix = {arXiv},
       eprint = {2607.06267},
 primaryClass = {cond-mat.quant-gas},
       adsurl = {https://ui.adsabs.harvard.edu/abs/2026arXiv260706267B}
}

@article{wuAdiabatic2012,
  title = {Adiabatic continuity between Hofstadter and Chern insulator states},
  author = {Wu, Ying-Hai and Jain, J. K. and Sun, Kai},
  journal = {Phys. Rev. B},
  volume = {86},
  issue = {16},
  pages = {165129},
  numpages = {13},
  year = {2012},
  month = {Oct},
  publisher = {American Physical Society},
  doi = {10.1103/PhysRevB.86.165129},
  url = {https://link.aps.org/doi/10.1103/PhysRevB.86.165129}
}

@article{bacciconi_prx2025_gravitonpolaritons,
  title = {Theory of Fractional Quantum Hall Liquids Coupled to Quantum Light and Emergent Graviton-Polaritons},
  author = {Bacciconi, Zeno and Xavier, Hernan B. and Carusotto, Iacopo and Chanda, Titas and Dalmonte, Marcello},
  journal = {Phys. Rev. X},
  volume = {15},
  issue = {2},
  pages = {021027},
  numpages = {35},
  year = {2025},
  month = {Apr},
  publisher = {American Physical Society},
  doi = {10.1103/PhysRevX.15.021027},
  url = {https://link.aps.org/doi/10.1103/PhysRevX.15.021027}
}

@ARTICLE{wangDynamics2025,
       author = {{Wang}, Yuzhu and {Huxford}, Joe and {Nguyen}, Dung Xuan and {Ji}, Guangyue and {Kim}, Yong Baek and {Yang}, Bo},
        title = "{Dynamics and lifetime of geometric excitations in moir{\'e} systems}",
      journal = {arXiv e-prints},
         year = 2025,
        month = feb,
          eid = {arXiv:2502.02640},
        pages = {arXiv:2502.02640},
          doi = {10.48550/arXiv.2502.02640},
archivePrefix = {arXiv},
       eprint = {2502.02640},
 primaryClass = {cond-mat.str-el},
       adsurl = {https://ui.adsabs.harvard.edu/abs/2025arXiv250202640W}
}

@ARTICLE{shenMagnetorotons2024,
       author = {{Shen}, Xiaoyang and {Wang}, Chonghao and {Hu}, Xiaodong and {Guo}, Ruiping and {Yao}, Hong and {Wang}, Chong and {Duan}, Wenhui and {Xu}, Yong},
        title = "{Magnetorotons in Moir{\'e} Fractional Chern Insulators}",
      journal = {arXiv e-prints},
         year = 2024,
        month = dec,
          eid = {arXiv:2412.01211},
        pages = {arXiv:2412.01211},
          doi = {10.48550/arXiv.2412.01211},
archivePrefix = {arXiv},
       eprint = {2412.01211},
 primaryClass = {cond-mat.str-el},
       adsurl = {https://ui.adsabs.harvard.edu/abs/2024arXiv241201211S}
}

@ARTICLE{paratera,
journal={Beijing PARATERA
Tech CO.,Ltd},
url = {https://cloud.paratera.com}
}

@article{xavier2025chiralgravitonslattice,
  title = {Chiral Graviton Modes on the Lattice},
  author = {Xavier, Hernan B. and Bacciconi, Zeno and Chanda, Titas and Son, Dam Thanh and Dalmonte, Marcello},
  journal = {Phys. Rev. Lett.},
  volume = {135},
  issue = {19},
  pages = {196501},
  numpages = {8},
  year = {2025},
  month = {Nov},
  publisher = {American Physical Society},
  doi = {10.1103/1636-kl65},
  url = {https://link.aps.org/doi/10.1103/1636-kl65}
}

@article{Hongyu_prl2024_thermodinamicFCI,
  title = {Thermodynamic Response and Neutral Excitations in Integer and Fractional Quantum Anomalous Hall States Emerging from Correlated Flat Bands},
  author = {Lu, Hongyu and Chen, Bin-Bin and Wu, Han-Qing and Sun, Kai and Meng, Zi Yang},
  journal = {Phys. Rev. Lett.},
  volume = {132},
  issue = {23},
  pages = {236502},
  numpages = {7},
  year = {2024},
  month = {Jun},
  publisher = {American Physical Society},
  doi = {10.1103/PhysRevLett.132.236502},
  url = {https://link.aps.org/doi/10.1103/PhysRevLett.132.236502}
}

@article{linExciton2022,
  title = {Exciton Proliferation and Fate of the Topological Mott Insulator in a Twisted Bilayer Graphene Lattice Model},
  author = {Lin, Xiyue and Chen, Bin-Bin and Li, Wei and Meng, Zi Yang and Shi, Tao},
  journal = {Phys. Rev. Lett.},
  volume = {128},
  issue = {15},
  pages = {157201},
  numpages = {8},
  year = {2022},
  month = {Apr},
  publisher = {American Physical Society},
  doi = {10.1103/PhysRevLett.128.157201},
  url = {https://link.aps.org/doi/10.1103/PhysRevLett.128.157201}
}

@article{KaiDasSarma_prl2011_flatbandsCB,
  title = {Nearly Flatbands with Nontrivial Topology},
  author = {Sun, Kai and Gu, Zhengcheng and Katsura, Hosho and Das Sarma, S.},
  journal = {Phys. Rev. Lett.},
  volume = {106},
  issue = {23},
  pages = {236803},
  numpages = {4},
  year = {2011},
  month = {Jun},
  publisher = {American Physical Society},
  doi = {10.1103/PhysRevLett.106.236803},
  url = {https://link.aps.org/doi/10.1103/PhysRevLett.106.236803}
}

@article{Sun2011_flat_chern_band,
	title = {Nearly Flatbands with Nontrivial Topology},
	author = {Sun, Kai and Gu, Zhengcheng and Katsura, Hosho and Das Sarma, S.},
	journal = {Phys. Rev. Lett.},
	volume = {106},
	issue = {23},
	pages = {236803},
	numpages = {4},
	year = {2011},
	month = {Jun},
	publisher = {American Physical Society},
	doi = {10.1103/PhysRevLett.106.236803},
	url = {https://link.aps.org/doi/10.1103/PhysRevLett.106.236803}
}

@article{Neupert2011_flat_chern_band,
	title = {Fractional Quantum Hall States at Zero Magnetic Field},
	author = {Neupert, Titus and Santos, Luiz and Chamon, Claudio and Mudry, Christopher},
	journal = {Phys. Rev. Lett.},
	volume = {106},
	issue = {23},
	pages = {236804},
	numpages = {4},
	year = {2011},
	month = {Jun},
	publisher = {American Physical Society},
	doi = {10.1103/PhysRevLett.106.236804},
	url = {https://link.aps.org/doi/10.1103/PhysRevLett.106.236804}
}

@article{Tang2011_flat_chern_band,
	title = {High-Temperature Fractional Quantum Hall States},
	author = {Tang, Evelyn and Mei, Jia-Wei and Wen, Xiao-Gang},
	journal = {Phys. Rev. Lett.},
	volume = {106},
	issue = {23},
	pages = {236802},
	numpages = {4},
	year = {2011},
	month = {Jun},
	publisher = {American Physical Society},
	doi = {10.1103/PhysRevLett.106.236802},
	url = {https://link.aps.org/doi/10.1103/PhysRevLett.106.236802}
}

@article{Regnault2011_FCI,
	title = {Fractional Chern Insulator},
	author = {Regnault, N. and Bernevig, B. Andrei},
	journal = {Phys. Rev. X},
	volume = {1},
	issue = {2},
	pages = {021014},
	numpages = {14},
	year = {2011},
	month = {Dec},
	publisher = {American Physical Society},
	doi = {10.1103/PhysRevX.1.021014},
	url = {https://link.aps.org/doi/10.1103/PhysRevX.1.021014}
}

@article{Sheng2011_FQAH_checkerboard_fermion,
	title = {Fractional quantum Hall effect in the absence of Landau levels},
	author = {Sheng, D.N.
	and Gu, Zheng-Cheng
	and Sun, Kai
	and Sheng, L.},
	journal = {Nature Communications},
	volume = {2},
	issue = {1},
	pages = {389},
	year = {2011},
	month = {Jul},
	doi = {10.1038/ncomms1380},
	url = {https://doi.org/10.1038/ncomms1380}
}

@article{Cai2023_signature_fqah_mote2,
	title = {Signatures of fractional quantum anomalous Hall states in twisted MoTe2},
	author = {Cai, Jiaqi
	and Anderson, Eric
	and Wang, Chong
	and Zhang, Xiaowei
	and Liu, Xiaoyu
	and Holtzmann, William
	and Zhang, Yinong
	and Fan, Fengren
	and Taniguchi, Takashi
	and Watanabe, Kenji
	and Ran, Ying
	and Cao, Ting
	and Fu, Liang
	and Xiao, Di
	and Yao, Wang
	and Xu, Xiaodong},
	journal = {Nature},
	volume = {622},
	issue = {7981},
	pages = {63-68},
	year = {2023},
	month = {Oct},
	doi = {10.1038/s41586-023-06289-w},
	url = {https://doi.org/10.1038/s41586-023-06289-w}
}

@article{Park2023_observation_fqah_mote2,
	title = {Observation of fractionally quantized anomalous Hall effect},
	author = {Park, Heonjoon
	and Cai, Jiaqi
	and Anderson, Eric
	and Zhang, Yinong
	and Zhu, Jiayi
	and Liu, Xiaoyu
	and Wang, Chong
	and Holtzmann, William
	and Hu, Chaowei
	and Liu, Zhaoyu
	and Taniguchi, Takashi
	and Watanabe, Kenji
	and Chu, Jiun-Haw
	and Cao, Ting
	and Fu, Liang
	and Yao, Wang
	and Chang, Cui-Zu
	and Cobden, David
	and Xiao, Di
	and Xu, Xiaodong},
	journal = {Nature},
	volume = {622},
	issue = {7981},
	pages = {74-79},
	year = {2023},
	month = {Oct},
	doi = {10.1038/s41586-023-06536-0},
	url = {https://doi.org/10.1038/s41586-023-06536-0}
}

@article{Zeng2023_thermo_evidence_fqah_mote2,
	title = {Thermodynamic evidence of fractional Chern insulator in moir\'e MoTe$_2$},
	author = {Zeng, Yihang
	and Xia, Zhengchao
	and Kang, Kaifei
	and Zhu, Jiacheng
	and Kn\"uppel, Patrick
	and Vaswani, Chirag
	and Watanabe, Kenji
	and Taniguchi, Takashi
	and Mak, Kin Fai
	and Shan, Jie},
	journal = {Nature},
	volume = {622},
	issue = {7981},
	pages = {69-73},
	year = {2023},
	month = {Oct},
	doi = {10.1038/s41586-023-06452-3},
	url = {https://doi.org/10.1038/s41586-023-06452-3}
}

@article{Xu2023_Observation_FQAH_tMote2,
	title = {Observation of Integer and Fractional Quantum Anomalous Hall Effects in Twisted Bilayer ${\mathrm{MoTe}}_{2}$},
	author = {Xu, Fan and Sun, Zheng and Jia, Tongtong and Liu, Chang and Xu, Cheng and Li, Chushan and Gu, Yu and Watanabe, Kenji and Taniguchi, Takashi and Tong, Bingbing and Jia, Jinfeng and Shi, Zhiwen and Jiang, Shengwei and Zhang, Yang and Liu, Xiaoxue and Li, Tingxin},
	journal = {Phys. Rev. X},
	volume = {13},
	issue = {3},
	pages = {031037},
	numpages = {12},
	year = {2023},
	month = {Sep},
	publisher = {American Physical Society},
	doi = {10.1103/PhysRevX.13.031037},
	url = {https://link.aps.org/doi/10.1103/PhysRevX.13.031037}
}

@article{Lu2024_FQAH_multilayer_graphene,
	title = {Fractional quantum anomalous Hall effect in multilayer graphene},
	author = {Lu, Zhengguang
	and Han, Tonghang
	and Yao, Yuxuan
	and Reddy, Aidan P.
	and Yang, Jixiang
	and Seo, Junseok
	and Watanabe, Kenji
	and Taniguchi, Takashi
	and Fu, Liang
	and Ju, Long},
	journal = {Nature},
	volume = {626},
	issue = {8000},
	pages = {759-764},
	year = {2024},
	month = {Feb},
	doi = {10.1038/s41586-023-07010-7},
	url = {https://doi.org/10.1038/s41586-023-07010-7}
}

@article{wang2023Geometric,
  title = {Geometric fluctuation of conformal Hilbert spaces and multiple graviton modes in fractional quantum Hall effect},
  volume = {14},
  ISSN = {2041-1723},
  url = {http://dx.doi.org/10.1038/s41467-023-38036-0},
  DOI = {10.1038/s41467-023-38036-0},
  number = {1},
  journal = {Nature Communications},
  publisher = {Springer Science and Business Media LLC},
  author = {Yuzhu,  Wang and Bo,  Yang},
  year = {2023},
  month = {apr} 
}

@article{hpc2021,
journal={HPC2021, Information Technology Services, The University of Hong Kong},
url={https://hpc.hku.hk/hpc/hpc2021/}}

@article{Cangemi2019Topological,
  title = {Topological Quantum Transition Driven by Charge-Phonon Coupling in the Haldane Chern Insulator},
  author = {Cangemi, L. M. and Mishchenko, A. S. and Nagaosa, N. and Cataudella, V. and De Filippis, G.},
  journal = {Phys. Rev. Lett.},
  volume = {123},
  issue = {4},
  pages = {046401},
  numpages = {6},
  year = {2019},
  month = {Jul},
  publisher = {American Physical Society},
  doi = {10.1103/PhysRevLett.123.046401},
  url = {https://link.aps.org/doi/10.1103/PhysRevLett.123.046401}
}

@article{Zezhu2026HaldaneHolstein,
       author = {{Wei}, Zezhu and {Wu}, Ang-Kun and {Xiao}, Di and {Lin}, Shi-Zeng},
        title = "{Haldane-Holstein model at fractional filling: Route to bosonic fractional Chern insulator and quantum anomalous Hall crystal}",
      journal = {arXiv e-prints},
         year = 2026,
        month = sep,
          eid = {arXiv:2609.14191},
        pages = {arXiv:2609.14191},
          doi = {10.48550/arXiv.2609.14191},
archivePrefix = {arXiv},
       eprint = {2609.14191},
 primaryClass = {cond-mat.str-el},
       adsurl = {https://ui.adsabs.harvard.edu/abs/2026arXiv260914191W}
}

@article{SousaJnior2026Realspace,
  title = {Real-Space Topology and Charge Order in the Haldane-Holstein Model},
  author = {Sousa-J\'unior, Sebasti\~ao dos A. and Fa\'undez, Juli\'an and Cysne, Tarik P. and Scalettar, Richard T. and Mondaini, Rubem},
  journal = {Phys. Rev. Lett.},
  volume = {136},
  issue = {25},
  pages = {256501},
  numpages = {8},
  year = {2026},
  month = {Jun},
  publisher = {American Physical Society},
  doi = {10.1103/s1my-sryj},
  url = {https://link.aps.org/doi/10.1103/s1my-sryj}
}

@article{Jeckelmann1998Density,
  title = {Density-matrix renormalization-group study of the polaron problem in the Holstein model},
  volume = {57},
  ISSN = {1095-3795},
  url = {http://dx.doi.org/10.1103/PhysRevB.57.6376},
  DOI = {10.1103/physrevb.57.6376},
  number = {11},
  journal = {Physical Review B},
  publisher = {American Physical Society (APS)},
  author = {Jeckelmann,  Eric and White,  Steven R.},
  year = {1998},
  month = Mar,
  pages = {6376–6385}
}

@article{Zhang1998Density,
  title = {Density Matrix Approach to Local Hilbert Space Reduction},
  volume = {80},
  ISSN = {1079-7114},
  url = {http://dx.doi.org/10.1103/PhysRevLett.80.2661},
  DOI = {10.1103/physrevlett.80.2661},
  number = {12},
  journal = {Physical Review Letters},
  publisher = {American Physical Society (APS)},
  author = {Zhang,  Chunli and Jeckelmann,  Eric and White,  Steven R.},
  year = {1998},
  month = Mar,
  pages = {2661–2664}
}

@article{Brockt2015Matric,
  title = {Matrix-product-state method with a dynamical local basis optimization for bosonic systems out of equilibrium},
  author = {Brockt, C. and Dorfner, F. and Vidmar, L. and Heidrich-Meisner, F. and Jeckelmann, E.},
  journal = {Phys. Rev. B},
  volume = {92},
  issue = {24},
  pages = {241106(R)},
  numpages = {5},
  year = {2015},
  month = {Dec},
  publisher = {American Physical Society},
  doi = {10.1103/PhysRevB.92.241106},
  url = {https://link.aps.org/doi/10.1103/PhysRevB.92.241106}
}

@article{Bauernfeind2020Time,
  title = {Time dependent variational principle for tree Tensor Networks},
  volume = {8},
  ISSN = {2542-4653},
  url = {http://dx.doi.org/10.21468/SciPostPhys.8.2.024},
  DOI = {10.21468/scipostphys.8.2.024},
  number = {2},
  journal = {SciPost Physics},
  publisher = {Stichting SciPost},
  author = {Bauernfeind,  Daniel and Aichhorn,  Markus},
  year = {2020},
  month = Feb 
}

@article{smcgBenedict1999,
  author  = {Benedict, Keith A. and Hills, R. K. and Mellor, C. J.},
  title   = {Theory of phonon spectroscopy in the fractional quantum {Hall} regime},
  journal = {Phys. Rev. B},
  volume  = {60},
  number  = {15},
  pages   = {10984--10996},
  year    = {1999},
  doi     = {10.1103/PhysRevB.60.10984}
}

@article{smcgBravyi2011,
  author  = {Bravyi, Sergey and DiVincenzo, David P. and Loss, Daniel},
  title   = {{Schrieffer-Wolff} transformation for quantum many-body systems},
  journal = {Ann. Phys.},
  volume  = {326},
  number  = {10},
  pages   = {2793--2826},
  year    = {2011},
  doi     = {10.1016/j.aop.2011.06.004}
}

\appendix
\newpage\clearpage
\renewcommand{\theequation}{S\arabic{equation}} \renewcommand{\thefigure}{S%
	\arabic{figure}} \setcounter{equation}{0} \setcounter{figure}{0}

\begin{widetext}
\begin{center}
    \textbf{\Large Supplemental Material for}\\[0.5em]
    \textbf{\Large"Spectroscopy of phonon-coupled integer and fractional Chern insulator: emergence of polarons and chirality deficit of graviton mode"}\\[1em]
\end{center}
In this supplemental material, we provide the Lang-Firsov transformation in the strong-coupling limit, Tree tensor network optimization and time evolution, endurance of Hall conductivity against electron-phonon coupling, convergence check, spectrum on all allowed momentum paths, finite-size extrapolation, effect of phonon frequency, and details on the perturbative treatment of electron-phonon interaction in Landau Level.
\section*{Strong coupling limit and Lang-Firsov transformation}
In this section, we provide a brief introduction to the Lang-Firsov transformation in the strong coupling limit(i.e., large $g$ limit)~\cite{mahan1981many}. It will be shown that the electron-phonon coupling will eventually localize the electrons, therefore giving us intuition about the formation of exciton-polarons and magnetoroton-polarons.

The idea is that, in this limit, the electron-phonon coupling term $g(n-\bar{n})(b^\dagger + b)$ dominates over the kinetic term and the electron-electron interaction term. Therefore, with only remain coupling term, the ground state would be the eigenstate of the phonon creation and annihilation operators, and at the same time, the eigenstate of the density operator, where the electron and phonon degrees of freedom are decoupled. Thus, CDW order will emerge to account for the electron degrees of freedom, while the phonon part forms a coherent state. Therefore, a transformation of phonon degrees of freedom from the occupation basis to the coherent basis could be made to correctly describe this limit, which is the Lang-Firsov transformation. 

 We start from the electronic Hamiltonian coupled to local Einstein phonons,
\begin{equation}
H=H_{\rm kin}+H_{\rm int}+H_{\rm ph}+H_{\rm ep},
\end{equation}
where
\begin{equation}
H_{\rm ph} = 
\omega_0\sum_i b_i^\dagger b_i,~H_{\rm ep} = g\omega_0 \sum_i
\left(b_i+b_i^\dagger\right)
\left(n_i-\bar n\right),~ \\ H_{\rm kin} =\sum_{ij}\left(t_{ij}c_i^\dagger c_j+\mathrm{H.c.}\right),~ H_{\rm int} = \sum_{ij}V_{ij}n_i n_j.
\end{equation}

The Lang--Firsov transformation is defined by the unitary operator
\begin{equation}
\widetilde H = 
e^{S}He^{-S},
\end{equation}
with
\begin{equation}
S=
g\sum_i \left(n_i-\bar n\right)\left(b_i^\dagger-b_i\right).
\end{equation}
The purpose of this transformation is to shift the equilibrium position of the local phonon oscillator depending on the electronic occupation. Using the Baker-Campbell-Hausdorff expansion, one finds
\begin{equation}
e^S b_i e^{-S} = 
b_i-g(n_i-\bar n),~e^S b_i^\dagger e^{-S} = 
b_i^\dagger-g(n_i-\bar n),~e^S n_i e^{-S}=n_i.
\end{equation}
As a consequence, the explicit linear electron--phonon coupling is exactly eliminated from the transformed Hamiltonian. Combining $H_{\rm ph}$ and $H_{\rm ep}$ gives
\begin{equation}
\widetilde H_{\rm ph}+
\widetilde H_{\rm ep} = \omega_0\sum_i b_i^\dagger b_i +
g^2\omega_0\sum_i\left(n_i-\bar n\right)^2.
\end{equation}
The second term is the familiar polaronic energy shift generated by the local lattice deformation.

The fermionic operators, on the other hand, acquire phonon displacement operators under the transformation,
\begin{equation}
e^S c_i e^{-S} = c_i X_i,~e^S c_i^\dagger e^{-S} = c_i^\dagger X_i^\dagger,
\end{equation}
where
\begin{equation}
X_i = 
\exp\left[-g\left(b_i^\dagger-b_i\right)\right].
\end{equation}
Therefore, every electronic hopping process is accompanied by a phonon cloud,
\begin{equation}
c_i^\dagger c_j
\longrightarrow
c_i^\dagger c_j
X_i^\dagger X_j,
\end{equation}
such that
\begin{equation}
\widetilde H_{\rm kin} = 
-\sum_{ij}\left[t_{ij}c_i^\dagger c_jX_i^\dagger X_j+\mathrm{H.c.}
\right].
\end{equation}
This reduction of the overlap between the initial and final phonon configurations is responsible for the polaronic suppression of the electronic kinetic energy.

To obtain a purely electronic effective Hamiltonian, we further treat the phonon displacement operators at the mean-field level by replacing them with their thermal expectation value,
\begin{equation}
X_i^\dagger X_j\rightarrow\left\langle X_i^\dagger X_j \right\rangle_{\rm ph}.
\end{equation}
For independent local Einstein phonons and ($i\neq j$), at zero temperature,
\begin{equation}
\left\langle
X_i^\dagger X_j
\right\rangle_{\rm ph}
= e^{-g^2},~t_{ij}^{\rm eff} = 
t_{ij}e^{-g^2}.
\end{equation}

Thus, all hopping amplitudes are reduced by the same factor within this local Holstein mean-field approximation. 

In contrast to the hopping operators, the density operator is invariant under the Lang--Firsov transformation.
Consequently, a bare density--density interaction is not directly dressed by the phonon displacement operators,
Nevertheless, its strength relative to the renormalized electronic kinetic energy is exponentially enhanced. 
 
 Summarizing, after the Lang-Firsov transformation and the mean-field transformation, the Hamiltonian becomes
 \begin{equation}
 \boxed{
     \begin{aligned}
         &H_{\rm kin} \rightarrow \widetilde H_{\rm kin} =  
-\sum_{ij}\left[t_{ij}e^{-g^2}c_i^\dagger c_j+\mathrm{H.c.}
\right]. \\ &
 H_{\rm ph} +  H_{\rm ep}\rightarrow\widetilde H_{\rm ph} + \widetilde H_{\rm ep} = \omega_0\sum_i b_i^\dagger b_i +
g^2\omega_0\sum_i\left(n_i-\bar n\right)^2. \\ & H_{\rm int} 
\rightarrow\widetilde H_{\rm int} = \sum_{ij}V_{ij}n_i n_j
     \end{aligned}}
 \end{equation}
 Defining the effective dimensionless interaction strength relative to the renormalized hopping gives, at zero temperature,
\begin{equation}
\frac{V_{ij}}{t_{ij}^{\rm eff}} = 
\frac{V_{ij}}{t_{ij}}
e^{g^2}.
\end{equation}
 In this sense, increasing the Holstein coupling drives the electronic system toward a more strongly correlated regime by exponentially suppressing charge motion while leaving the bare repulsive interaction intact. In addition to the change of energy scale, the electron-phonon coupling tends to localize electrons.
 
\section*{Tree Tensor Network, Optimization and Time Evolution}
Historically, two strategies have been proposed to reduce the dimensionality of the phonon Hilbert space in the context of the one-dimensional Holstein problem and the coarse-grained (traditional) version of the DMRG algorithm. The pseudo-spin approach encodes the phonon degrees of freedom by adding extra pseudo-sites to the 1D chain~\cite{Jeckelmann1998Density}, while the Local Basis Optimization (LBO) method variationally truncates the local phonon Hilbert space at every DMRG sweep by introducing an isometric tensor at every site~\cite{Zhang1998Density, Brockt2015Matric}.

The method adopted in this work follows the spirit of LBO; actually, a combination of LBO and the MPS description on a quantum lattice gives a \textit{tree tensor network} on which we can use the idea of tangent space~\cite{Haegeman2011Time} to perform time evolution~\cite{Bauernfeind2020Time}. The detailed derivation of tangent space and time evolution for general tree tensor networks can be found in ~\cite{Bauernfeind2020Time}, In this section, we briefly explain the tree tensor network(TNN) structure for the Holstein problem, ground state optimization, and time evolution. We emphasize the nomenclature of tree tensor network, rather than MPS + LBO, because we want to highlight the difference in the underlying tangent-space structure.
\begin{figure}[h!]
    \centering
    \includegraphics[width=0.4\linewidth]{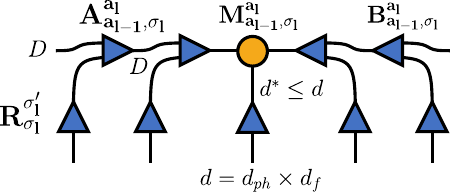}
    \caption{\textbf{Tree tensor network structure for phonon coupled with fermion} On every site a extra isometric tensor is introduced to compress phonon Hilbert space from $d = d_{ph}\times d_f$ to $d^*$, where $d_{ph} $ labels the dimension of the single site truncated phonon occupation basis and $d_f$ labels the dimension of single site fermionic occupation basis. Here $a_l$ labels the virtual bond while $\sigma_l$ labels the physical space.}
    \label{fig:supple_ttn}
\end{figure}

As mentioned before, an isometric tensor is introduced on every site to truncate the phonon Hilbert Space. The extra tensor, which we denote as the $R$ tensor in Fig.~\ref{fig:supple_ttn}, extracts the essential contribution from phonons by tracing out all the other tensors, diagonalizing the single-site phonon density matrix, and discarding the component with small weight. Although an extra $R$ tensor has been added, the nature of the TNN structure is still retained, meaning a canonical center could be defined, from which we can easily obtain the density matrix and extract the important component of the environment. Given that, the optimization of a TNN naturally follows the spirit of DMRG: one sweeps over all sites, projects out the environment, and moves the canonical center along with the optimization region.

The time evolution is performed in the tangent space of this TTN; the tangent space structure and projection of a TTN into it have been studied in ~\cite{Bauernfeind2020Time}, and the procedure is similar to TDVP on MPS.

\begin{figure}[h!]
    \centering
    \includegraphics[width=1\linewidth]{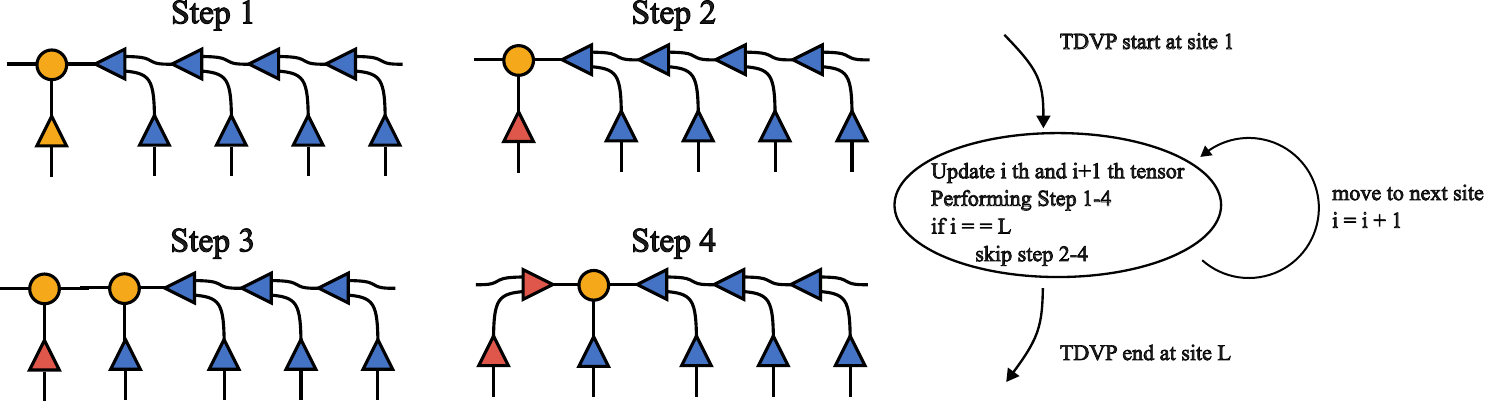}
    \caption{\textbf{TDVP sequence on this 1D tree tensor network}, Update sequence to perform two-site TDVP time step from
    time $t$ to $t + \Delta t$ for the TTN shown in Fig.~\ref{fig:supple_ttn} at site $1$.  Yellow denotes tensors that are updated in the current step. Red indicates the tensor has been taken at time $t+\Delta t$ and will not changed at current sweep; blue tensors are still taken at time $t$ and will be updated in  later sweep of the remaining sites. Triangles indicate the orthogonalization of each tensor. Updates on two-site tensors are in the forward direction (negative sign), while updates on a single site are backwards time
evolutions. }
    \label{fig:supple_ttn_tdvpsq}
\end{figure}

A full two-site TDVP step on TTN can be performed by(also illustrated in Fig.~\ref{fig:supple_ttn_tdvpsq}):
\begin{itemize}
    \item Start from site 1 and end at site L, where L is the length of the MPS chain.
    \item Loop site i from i = 1:L\begin{itemize}
        \item if j !=L, 
        \begin{itemize}
            \item Step 1: obtain the effective Hamiltonian $H_{eff}^j$ by tracing out all the other sites (except $R_j$), and update the tensor at site $\tilde{M}_j = R_jM_j$ as $\tilde{M}_j(t+\Delta t) = \exp(-i \Delta t H_{eff}^j)\tilde{M}_j(t)$, then split $\tilde{M}_{j}$ back to $M_j$ and $M_j$ by Singular Value decomposite(SVD)
            \item Step 2: obtain the 1 site effective Hamiltonian $H_{eff,1site}^j$ by further trace out $R_j$, and update $M_j$ as $M_j(t+\Delta t) = \exp(i \Delta t H_{eff,1site}^j)M_j(t)$
            \item Steps 3 and 4 are very similar to the 2-site TDVP scheme for MPS, with the effective Hamiltonian changed by further tracing out $R$ tensors on the related sites
        \end{itemize}
        \item if i == L, only perform Step 1
    \end{itemize}
    \item End
\end{itemize}

\section*{Endurance of Hall conductivity against electron-phonon coupling}
In this section, we study the response of the FCI phase to the external flux when the electron-phonon coupling is present, where we show that the Hall conductivity is robust and the topological properties of the FCI ground state are still preserved, thereby serving as the prerequisite for the magneto-roton and graviton mode study presented in the main text.

We also note that, for the integer-filling case, the robustness of the CI phase has been studied using the cluster perturbation method~\cite{Cangemi2019Topological} and quantum Monte Carlo~\cite{SousaJnior2026Realspace}.

By threading an external $2\pi$ flux along the cylinder, the charge response will be $\Delta n = \sigma \phi$ according to Laughlin's argument~\cite {laughlin1983anomalous}. Flux threading is implemented by imposing twisted boundary conditions along the y direction of the cylinder and adiabatically evolving the ground state. The charge pumped is measured by the charge accumulation on the edges. 

In Fig.~\ref{fig:supple_chargepumping} we show the charge pumping result for $g = 0.4$, the quantized value, and the entanglement spectrum flow shows the robustness of the FCI ground state with electron-phonon coupling. Combined with previous numerical work~\cite{Cangemi2019Topological,SousaJnior2026Realspace}, we demonstrate the robustness of Hall conductivity.

\begin{figure}[htp!]
    \centering
    \includegraphics[width=0.8\linewidth]{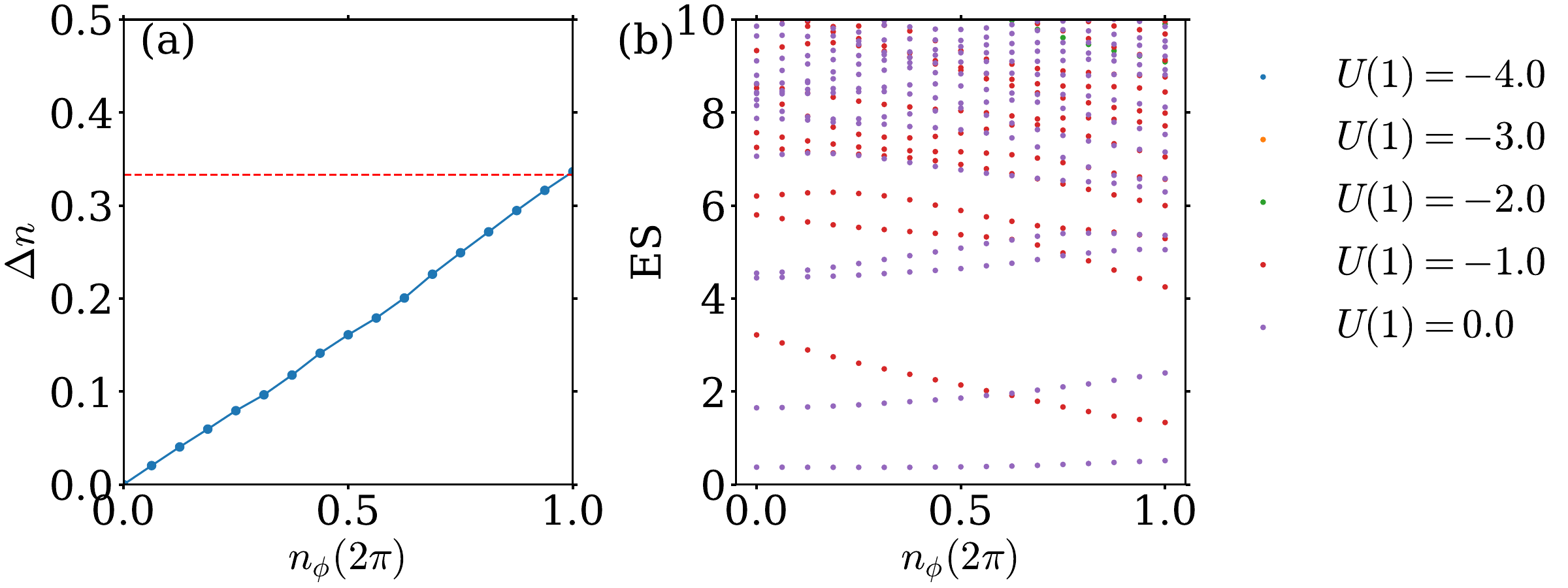}
    \caption{\textbf{Hall conductivity at $g = 0.4$}(a). By threading $2\pi$ flux along the cylinder, the $1/3$ FCI phase will pump  $1/3$ quantized charge from the left edge of the cylinder to the right edge. The red dashed line labels $\Delta n = 1/3$; (b) Entanglement spectrum flow against the external flux.}
    \label{fig:supple_chargepumping}
\end{figure}

\section*{Convergence Check}
In this section, we provide information about the convergence test of phonon Hilbert space truncation and bond dimension convergence in DMRG and TDVP.
\subsection*{Truncation on phonon space}
For the Hamiltonian with generic form
\begin{equation}
    H = -\sum_{ij}t_{i,j}c^\dagger_ic_j + V\sum_{ij}n_in_j + \omega\sum_i b^\dagger_i b_i +  g\omega_0 \sum_i (n_i - \bar{n})(b_i^\dagger + b_i)
\end{equation}
The Hamiltonian implies the local commutation relations
\begin{equation}
[H,b_i]=-\omega b_i-g\omega(n_i-\bar n),\qquad
[H,b_i^\dagger]=\omega b_i^\dagger+g\omega(n_i-\bar n).
\end{equation}
For an exact energy eigenstate, the expectation value of each
commutator vanishes. Adding the resulting equations gives the
site-resolved identity
\begin{equation}
\langle b_i+b_i^\dagger\rangle
=-2g\langle n_i-\bar n\rangle .
\label{eq:displacement_identity}
\end{equation}
We use this identity as a consistency check for the DMRG results
by evaluating the root-mean-square residual
\begin{equation}
\epsilon_{\mathrm{disp}}
=
\left[
\frac{1}{N}\sum_i
\left|
\langle b_i+b_i^\dagger\rangle
+2g\langle n_i-\bar n\rangle
\right|^2
\right]^{1/2}.
\label{eq:displacement_residual}
\end{equation}
A small $\epsilon_{\mathrm{disp}}$ indicates that the independently
computed charge density and phonon displacement satisfy the
exact eigenstate relation.
\begin{figure}[t]
    \centering
    \includegraphics[width=0.5\columnwidth]{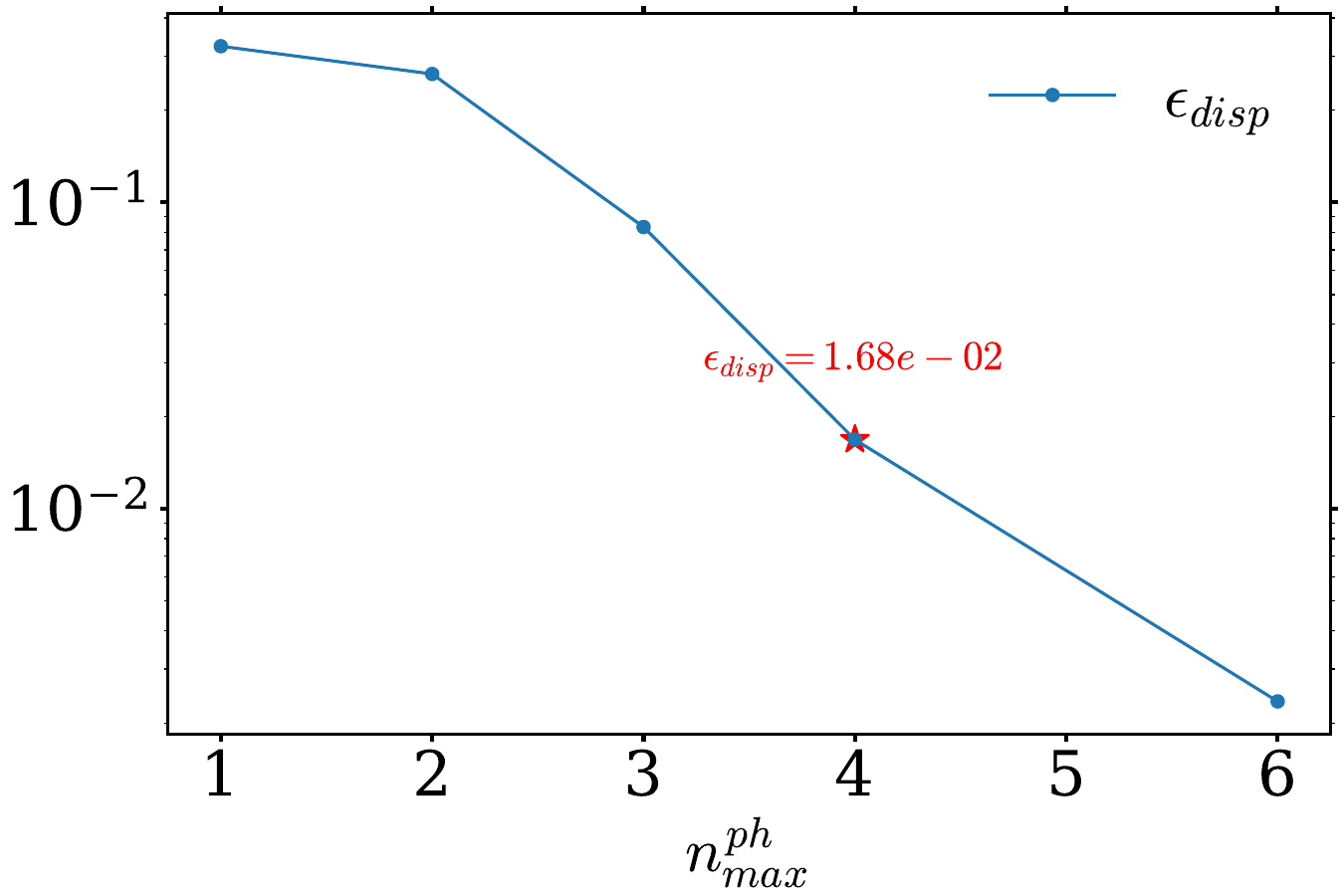}
    \caption{Displacement-identity residual $\epsilon_{\mathrm{disp}}$
    as a function of the maximum local phonon occupation
    $n_{\max}^{\mathrm{ph}}$. The red star marks
    $n_{\max}^{\mathrm{ph}}=4$ which we used in the main text with
    $\epsilon_{\mathrm{disp}}=1.68\times10^{-2}$.}
    \label{fig:displacement_convergence}
\end{figure}

As shown in Fig.~\ref{fig:displacement_convergence},
$\epsilon_{\mathrm{disp}}$ decreases systematically as the local
phonon cutoff is increased. At the cutoff
$n_{\max}^{\mathrm{ph}}=4$ used in our calculations, the residual
is $1.68\times10^{-2}$; increasing the cutoff to
$n_{\max}^{\mathrm{ph}}=6$ reduces it further to approximately
$2\times10^{-3}$. The residual decreases systematically as the phonon space is
enlarged. At $n_{\max}^{\mathrm{ph}}=4$, used in our main
calculations, we obtain $\epsilon_{\mathrm{disp}}
=1.68\times10^{-2}$, indicating that the charge density and
phonon displacement satisfy the eigenstate relation to this
accuracy.

This result is performed on an $L_y\times L_x=3\times16$
cylinder at $\bar n=1/2$ filling in the CDW phase, where the phonon
occupation is higher than in the CI phase. It therefore provides
a more stringent check of the phonon cutoff; truncation effects
are expected to be smaller in the CI phase.

\subsection*{DMRG convergence}
The bond dimension $D$ is a key parameter governing the accuracy of DMRG calculations. Due to the large physical dimension $d = 2\times (n_{max}^{ph} + 1)$ employed in this simulation, the maximum value of $D$ we can use is limited. In this section, we examine the convergence of the results presented in the main text.

We examine the convergence of the ground state with respect to the
bond dimension $D$ using the bipartite entanglement entropy
$S_{\mathrm E}(x)$ across a cut at position $x$.

\begin{figure}[ht!]
    \centering
    \includegraphics[width=0.7\columnwidth]{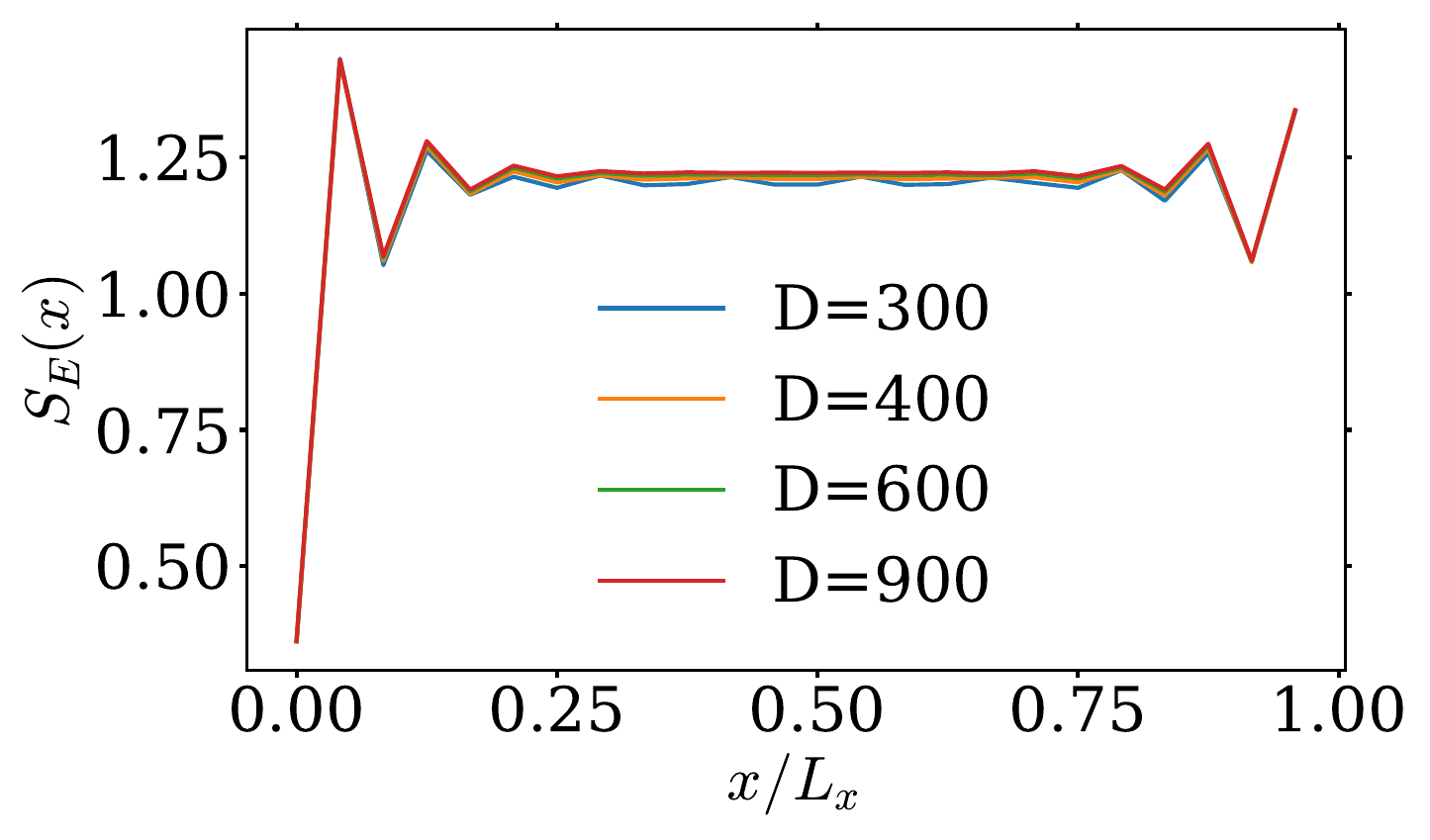}
    \caption{Entanglement entropy $S_{\mathrm E}(x)$ along the
    cylinder for different bond dimensions $D$. The curves for
    $D=300$, $400$, $600$, and $900$ nearly overlap throughout
    the system.}
    \label{fig:EE_bond_convergence}
\end{figure}

As shown in Fig.~\ref{fig:EE_bond_convergence}, increasing the
bond dimension from $D=300$ to $D=900$ produces only minor changes
in the entanglement-entropy profile. The agreement holds both in
the bulk and near the boundaries, supporting convergence of the
ground-state entanglement with respect to $D$ already at $D=300$.

This result is performed on an $L_y\times L_x=3\times24$
cylinder at $\bar n=1/6$ filling with $g = 1.6$, where the phonon fluctuation has strongly dressed the magnetoroton mode and forms polarons and serves as a more stringent check for the convergence of the DMRG calculation.
\subsection*{TDVP convergence}
We next examine the convergence of the time evolution used to extract dynamical correlations and spectra. The evolving state can develop substantially more entanglement than the ground state, making its convergence with bond dimension more demanding.
\subsubsection{Exciton-polarons}
\begin{figure}[ht!]
    \centering
    \includegraphics[width=0.5\columnwidth]{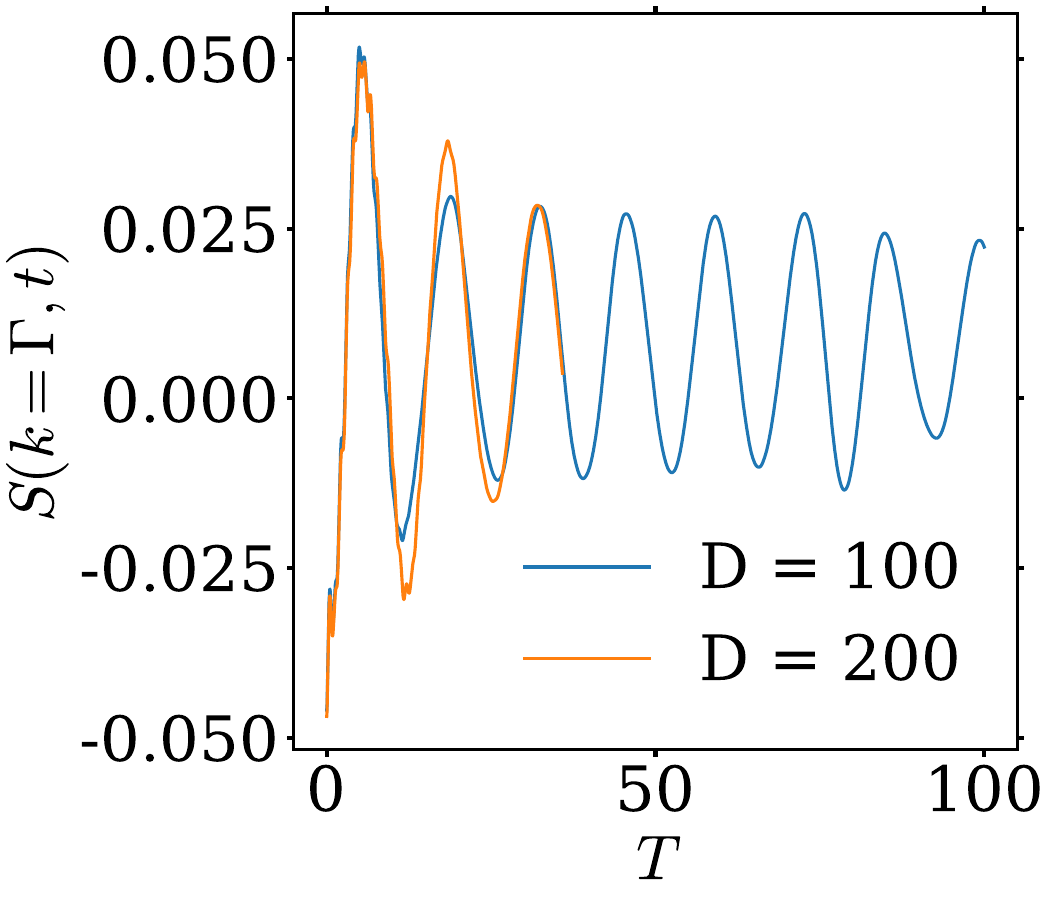}
    \caption{Bond-dimension convergence of the time-dependent
    density correlation $S(k=\Gamma,t)$ in the CI phase.
    The $\Gamma$ point corresponds to the minimum of the
    exciton-polaron dispersion. Results for $D=100$ and $D=200$
    show similar oscillation periods over their common time
    interval; the $D=200$ data extend to approximately $t=35$.}
    \label{fig:tdvp_exciton_convergence}
\end{figure}

We examine the bond-dimension convergence of the time evolution
at the $\Gamma$ point, where the exciton-polaron mode reaches its
energy minimum. As shown in
Fig.~\ref{fig:tdvp_exciton_convergence}, increasing $D$ from
$100$ to $200$ leaves the dominant oscillation period of
$S(\Gamma,t)$ nearly unchanged over the common time window.
This indicates that the energy of the low-lying
exciton-polaron mode is approximately converged with respect to
the bond dimension. 

\subsubsection{Magnetoroton-polarons}
\begin{figure}[ht!]
    \centering
    \includegraphics[width=0.5\columnwidth]{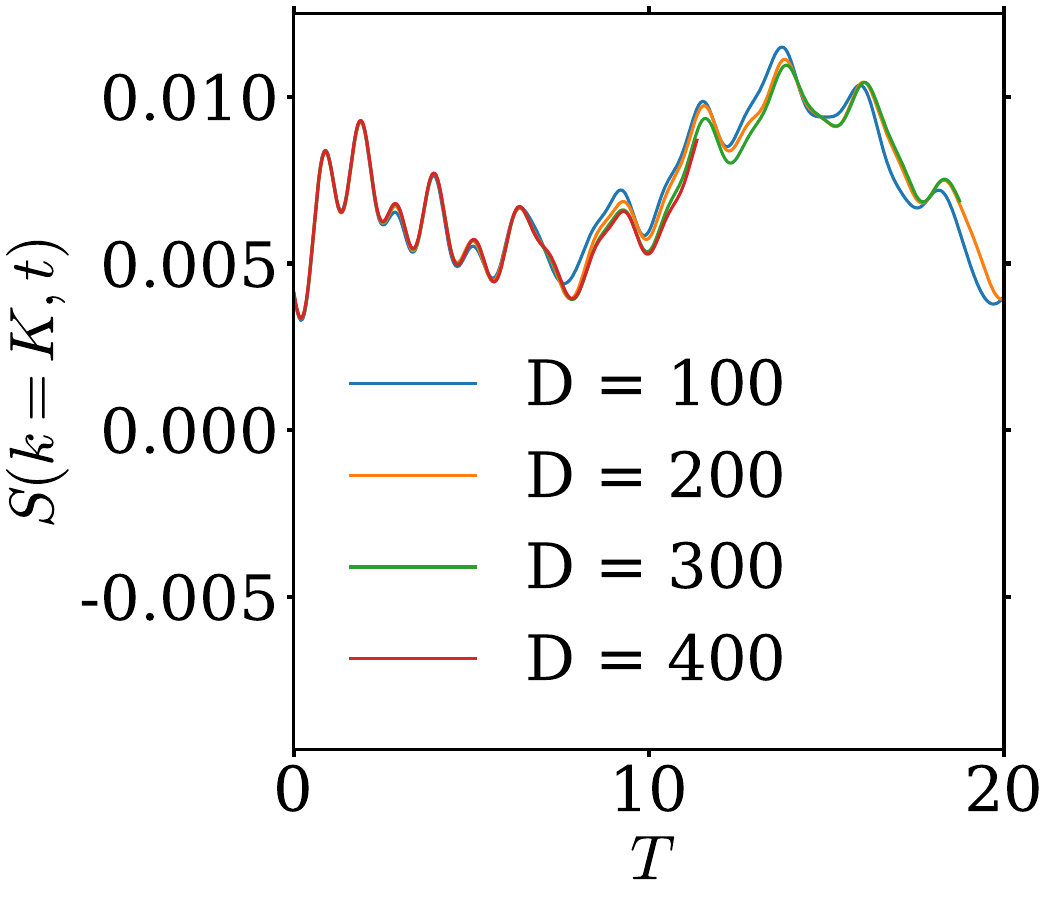}
    \caption{Bond-dimension convergence of the time-dependent
    density correlation $S(\mathbf{k}=K,t)$ in the FCI phase.
    The $K$ point corresponds to the magnetoroton minimum.
    Results for $D=100$, $200$, $300$, and $400$ agree closely
    over their common time interval.}
    \label{fig:tdvp_roton_convergence}
\end{figure}

For the fractional-filling case, we test the time evolution at
the $K$ point, where the magnetoroton reaches its energy minimum.
As shown in Fig.~\ref{fig:tdvp_roton_convergence}, the
time-dependent correlations obtained with different bond
dimensions exhibit closely matching oscillations over their
common time window. In particular, the $D=200$ and $D=300$
results remain close throughout the displayed evolution,
while the shorter $D=400$ trace provides an additional check.
This agreement supports convergence of the $K$-point
magnetoroton response with respect to the bond dimension.

\subsection{Chiral graviton mode}
\begin{figure*}[t]
    \centering
    \includegraphics[width=0.8\columnwidth]{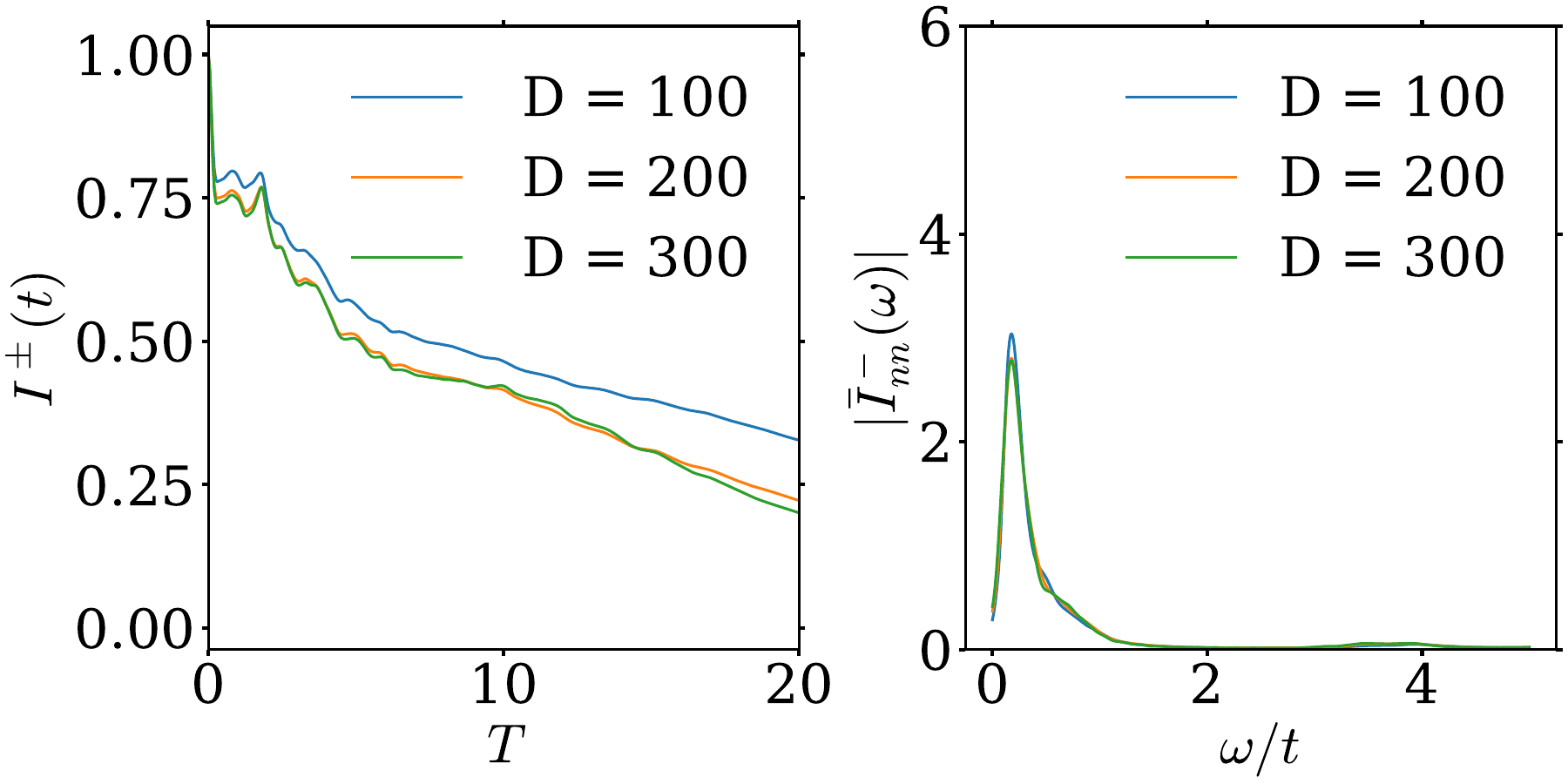}
    
    \caption{Bond-dimension convergence of the graviton response.
    Left: time-dependent correlation $|I^{\pm}(t)|$ for
    $D=100$, $200$, $300$. Right: the corresponding
    negative-helicity spectrum $|\bar I^-_{nn}(\omega)|$.
    The resonance position and spectral weight change little with increasing $D$ after $D = 200$.}
    \label{fig:tdvp_graviton_convergence}
\end{figure*}

We further examine the bond-dimension convergence of the graviton
response. As shown in Fig.~\ref{fig:tdvp_graviton_convergence},
the $D=100$ time trace deviates from the higher-$D$ results at
later times, whereas the curves for $D=200$ and $300$
remain close. In the frequency domain, increasing $D$ leaves both the height of the dominant negative-helicity peak
nearly unchanged. Thus, the graviton spectrum is stable against the bond-dimension increase examined
here.

\section*{Spectrum on all allowed momentum paths}
\subsection{Exciton-polaron}
\begin{figure}[ht!]
    \centering
    \includegraphics[width=\linewidth]{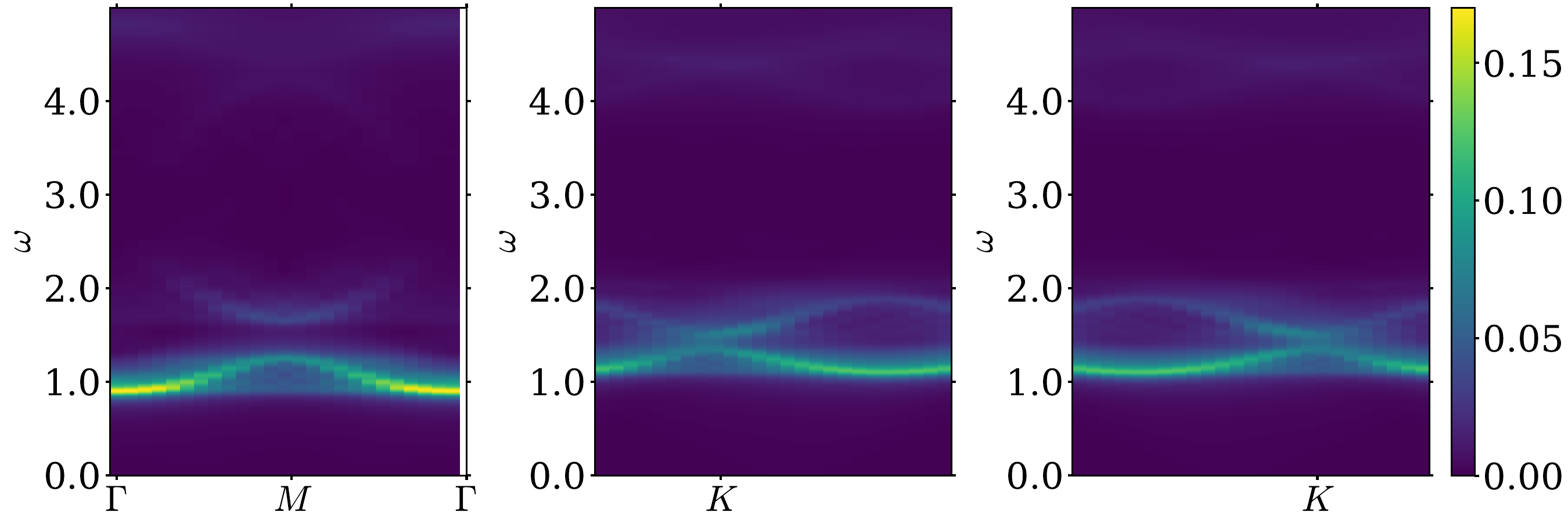}
    \caption{
    Momentum-resolved density-response spectrum in the CI at
    $g=1.2$. The three panels show the momentum cuts associated
    with the three transverse-momentum sectors accessible on the
    $L_y=3$ cylinder. The first cut follows
    $\Gamma$--$M$--$\Gamma$, while the other two pass through the
    Brillouin-zone corners. The dominant low-energy branch reaches
    its minimum at $\Gamma$, confirming that the lowest excitation
    is the $\Gamma$-centered exciton-polaron mode.
    }
    \label{fig:exciton_fullpath}
\end{figure}
To determine the momentum dependence of the exciton-polaron mode, we calculate the density-response spectrum along all momentum cuts accessible on the $L_y=3$ cylinder. As shown in Fig.~\ref{fig:exciton_fullpath}, a well-defined low-energy branch persists throughout the momentum space. Its energy reaches its minimum at the $\Gamma$ point and increases toward the $M$ point and the other transverse-momentum sectors. This confirms that the $\Gamma$-point response discussed in the main text captures the lowest exciton-polaron excitation. A weaker higher-energy branch is also visible, but remains separated from the dominant low-energy mode over most of the accessible momenta. We also note that path-2 and path-3 are related by spatial inversion.

\subsection{Magnetoroton-polaron}
\begin{figure}[ht!]
    \centering
    \includegraphics[width=\linewidth]{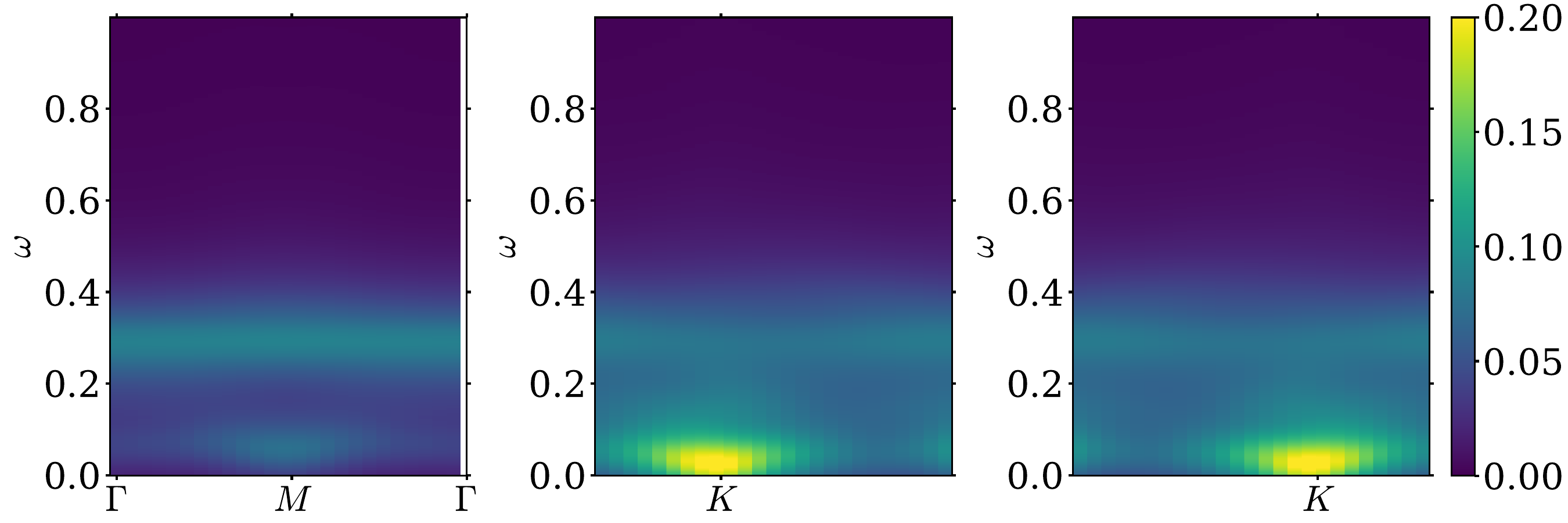}
    \caption{
    Momentum-resolved density-response spectrum
    $S(\mathbf{k},\omega)$ in the FCI at $g=1.8$. The three panels
    show the momentum cuts associated with the transverse-momentum
    sectors accessible on the $L_y=3$ cylinder. The first cut follows
    $\Gamma$--$M$--$\Gamma$, while the other two pass through the
    Brillouin-zone corners $K$. The lowest and most pronounced
    low-energy response occurs at $K$, identifying it as the minimum
    of the softened magnetoroton-polaron branch and connecting this
    excitation to the ordering wave vector of the ensuing CDW phase.
    }
    \label{fig:magnetoroton_fullpath}
\end{figure}
We further examine the magnetoroton dispersion along all momentum
cuts accessible on the $L_y=3$ cylinder. As shown in
Fig.~\ref{fig:magnetoroton_fullpath}, low-energy spectral weight is
present near both the $M$ and $K$ points, but the sharpest and lowest
mode occurs at the Brillouin-zone corners $K$. At $g=1.8$, this mode
has already softened to an energy well below the broad higher-energy
response around $\omega\simeq 0.3$. The concentration of low-energy
spectral weight at $K$ confirms that the magnetoroton-polaron minimum
is located at the ordering wave vector of the ensuing CDW phase.
Consequently, the $K$-point response discussed in the main text
captures the dominant low-energy instability of the FCI.
\section*{Finite size extrapolation}
\begin{figure}[ht!]
    \centering
    \includegraphics[width=\linewidth]{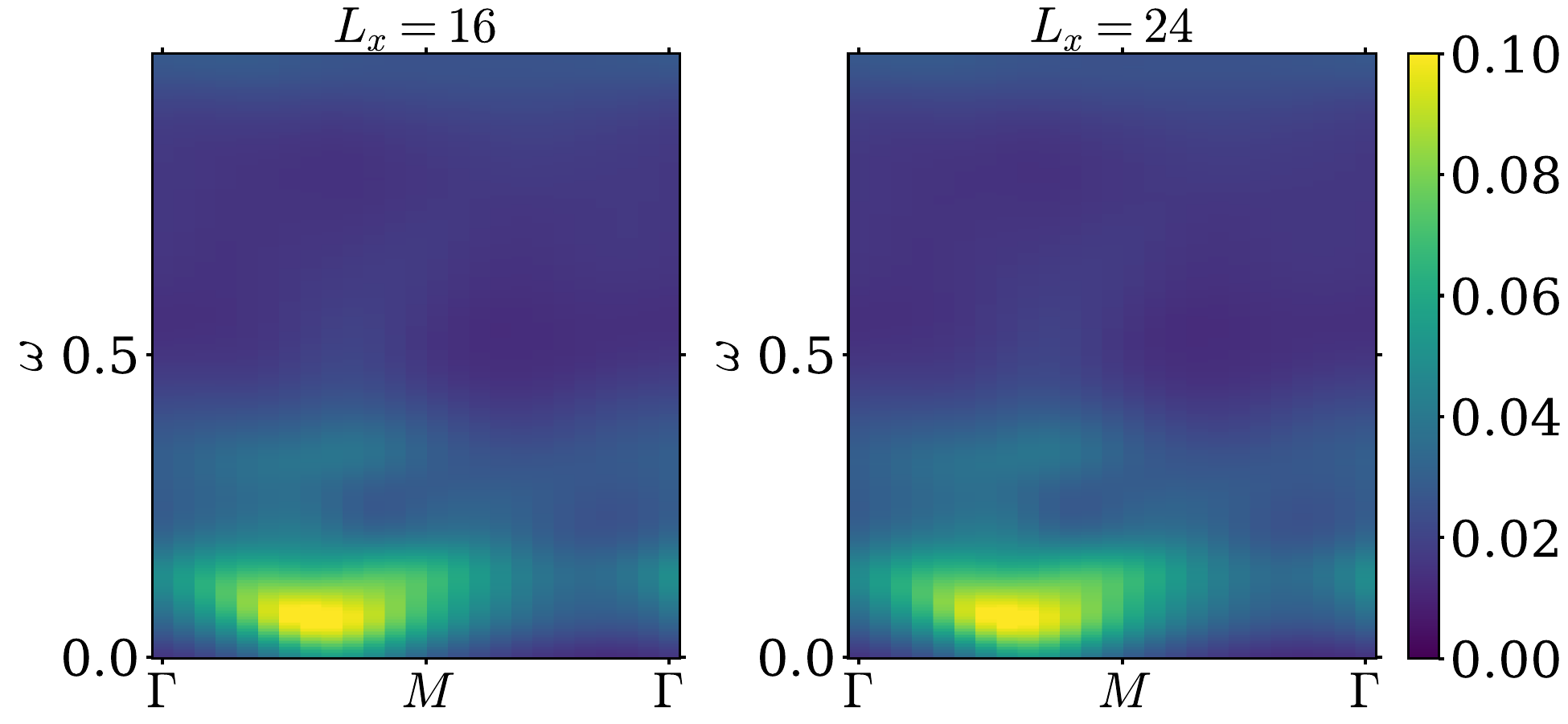}
    \caption{
    Dependence of the density-response spectrum
    $S(\mathbf{k},\omega)$ on the cylinder length. Results are shown
    along the momentum cut passing through $K$ for $L_x=16$ and
    $L_x=24$, with fixed circumference $L_y=3$ and coupling $g=1.2$.
    The low-energy magnetoroton branch and its minimum at $K$ remain
    essentially unchanged upon increasing $L_x$, indicating weak
    finite-length effects in the extracted dispersion.
    }
    \label{fig:length_dependence}
\end{figure}
To examine finite-length effects in the dynamical response, we compare
$S(\mathbf{k},\omega)$ on $L_y=3$ cylinders with $L_x=16$ and
$L_x=24$. As shown in Fig.~\ref{fig:length_dependence}, increasing
$L_x$ leaves the position and dispersion of the low-energy
magnetoroton mode essentially unchanged. In particular, the spectral
minimum remains located at $K$, with only minor changes in the
distribution and broadening of the spectral weight. The agreement
between the two lengths indicates that the magnetoroton dispersion
reported in the main text is not substantially affected by the finite
cylinder length.

We note that the next larger circumference retaining \(K\) as an allowed momentum is \(L_y=6\). Such a calculation is computationally demanding because the bond dimension required by DMRG grows exponentially with the cylinder circumference, further compounded by the enlarged local Hilbert space needed to accommodate phonons.

\section*{Effect of phonon frequency}
\subsection{Exciton-polarons}
\begin{figure}[ht!]
    \centering
    \includegraphics[width=0.5\linewidth]{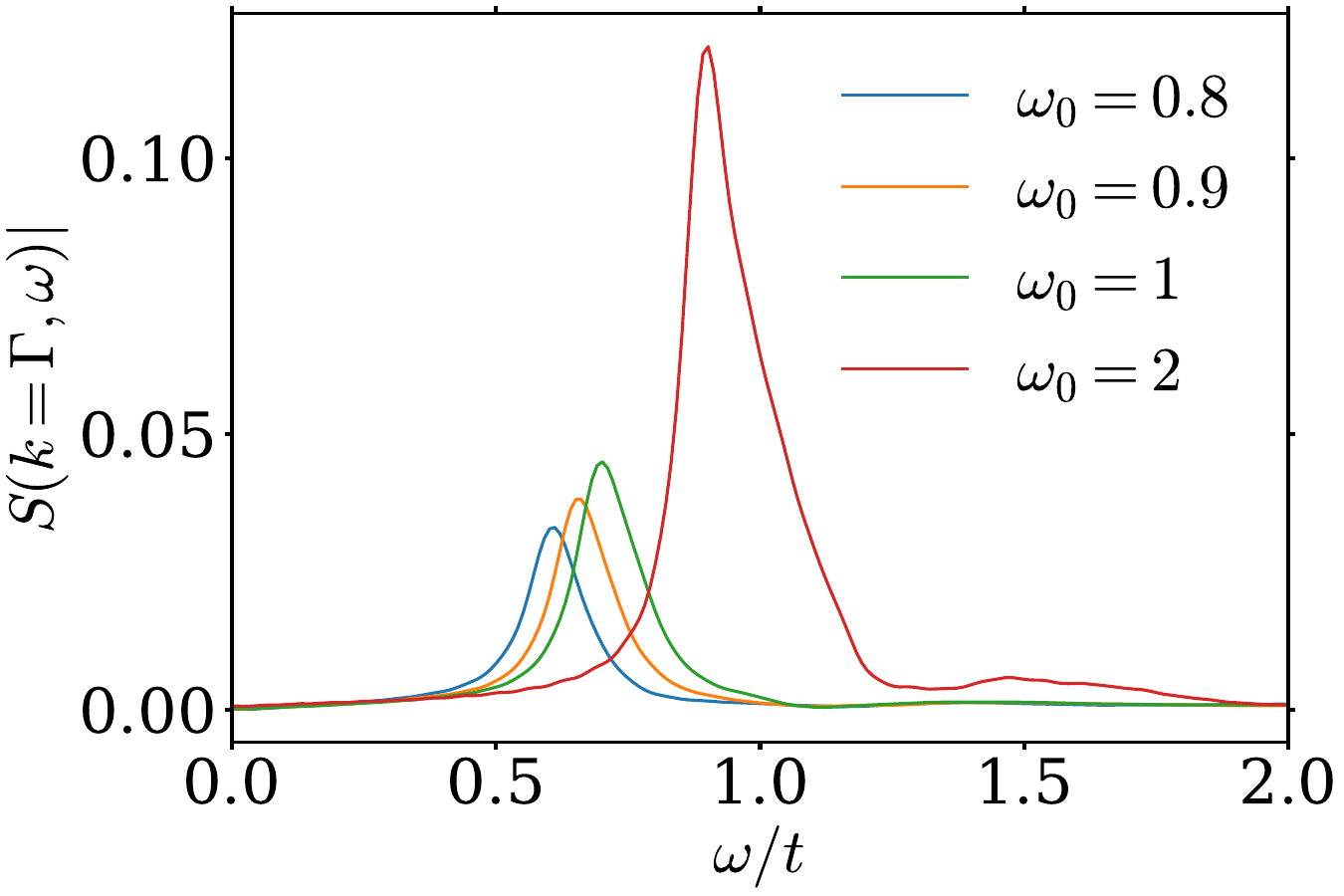}
    \caption{
    Density-response spectrum $S(\Gamma,\omega)$ of the CI for
    different bare phonon frequencies $\omega_0$ at fixed $g=1.2$.
    The dominant exciton-polaron peak shifts continuously with
    increasing $\omega_0$, while no level splitting or avoided
    crossing is resolved. Since the electron--phonon vertex is
    $g\omega_0$, this scan changes both the phonon frequency and the
    dimensional coupling strength.
    }
    \label{fig:exciton_tune_omega}
\end{figure}
To examine whether the exciton-polaron mode exhibits a resonant level
crossing with the bare phonon, we vary the phonon frequency $\omega_0$
across the exciton energy scale and calculate the density response
spectrum at $\Gamma$. As shown in Fig.~\ref{fig:exciton_tune_omega},
the dominant exciton-polaron peak shifts continuously toward higher
energy as $\omega_0$ increases. Within the available frequency
resolution, we observe neither a second low-energy branch nor a
resolvable avoided crossing. The evolution is therefore more
consistent with a continuous phonon-induced renormalization of the
exciton than with a simple two-level hybridization picture.

We note that the electron--phonon vertex in our convention is
$g\omega_0$. Consequently, varying $\omega_0$ at fixed $g$ changes
both the bare phonon frequency and the dimensional coupling strength.
The results should therefore be interpreted as the evolution of the
fully dressed excitation rather than as a pure detuning scan at fixed
hybridization.

\subsection{Magnetoroton polarons}
\begin{figure}[ht!]
    \centering
    \includegraphics[width=0.85\linewidth]{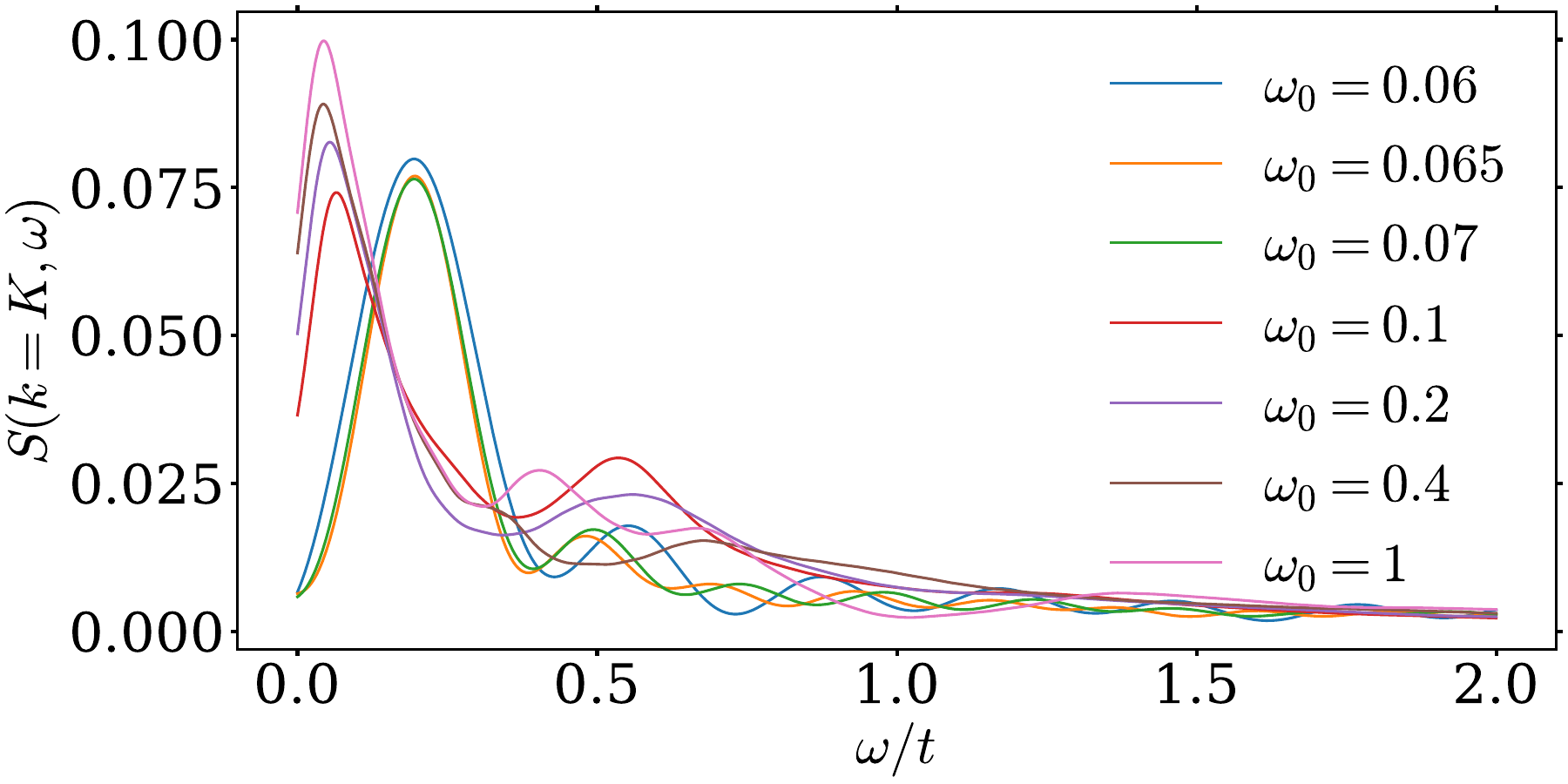}
    \caption{
    Density-response spectrum $S(K,\omega)$ of the FCI for different
    bare phonon frequencies $\omega_0$ at fixed $g=1.2$. As
    $\omega_0$ increases, spectral weight is transferred from the
    finite-energy magnetoroton peak near $\omega\simeq0.2$ to a
    strongly softened low-energy response. No resolved level crossing
    or avoided crossing is observed. The Fourier broadening is
    $\eta=0.05$.
    }
    \label{fig:magnetoroton_tune_omega}
\end{figure}
We next examine the dependence of the magnetoroton response on the
bare phonon frequency $\omega_0$. As shown in
Fig.~\ref{fig:magnetoroton_tune_omega}, for
$\omega_0=0.06$--$0.07$, the dominant magnetoroton peak remains near
$\omega\simeq 0.2$. Upon increasing $\omega_0$, spectral weight is
rapidly transferred to much lower frequencies, indicating a strong
softening of the magnetoroton mode. At the same time, broad
higher-energy features remain visible between approximately
$\omega=0.4$ and $0.8$. We do not observe two well-resolved branches
that undergo a level crossing or exchange their spectral character.
Instead, the spectra reveal a continuous phonon-induced
renormalization and redistribution of the many-body magnetoroton
response.

Because the electron--phonon vertex is $g\omega_0$ in our convention,
increasing $\omega_0$ at fixed $g$ simultaneously increases the
dimensional coupling strength. The observed softening therefore
reflects the combined effects of changing the phonon energy and
strengthening the electron--phonon interaction, rather than a pure
detuning between two modes. Moreover, since the broadening
$\eta=0.05$ is comparable to the lowest energy scales, the spectra do
not resolve whether the softened mode develops a strictly vanishing
gap.

\subsection{Chiral graviton mode}
\begin{figure}[ht!]
    \centering
    \includegraphics[width=0.75\linewidth]{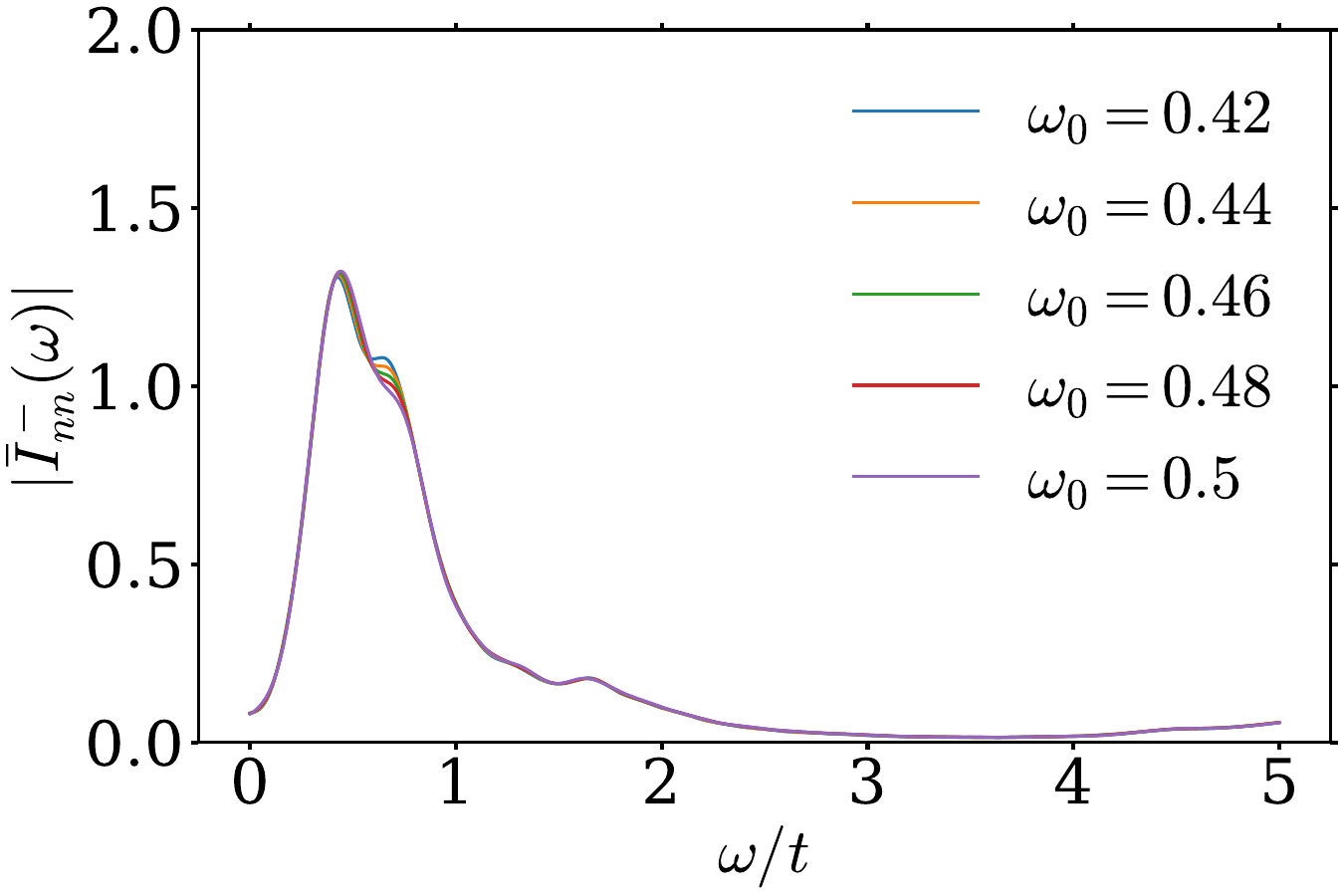}
    \caption{
    Normalized negative-chirality graviton response
    $|\bar{I}_{nn}^{-}(\omega)|$ for different bare phonon
    frequencies $\omega_0$ at fixed $g=0.4$. The dominant peak and
    the overall line shape remain nearly unchanged as $\omega_0$ is
    tuned from $0.42$ to $0.50$. No resolvable level splitting or
    avoided crossing is observed. The Fourier broadening is
    $\eta=0.05$.
    }
    \label{fig:graviton_tune_omega}
\end{figure}
Finally, we tune the bare phonon frequency across the characteristic
energy of the graviton response. Figure~\ref{fig:graviton_tune_omega}
shows the normalized spectrum in the negative-chirality channel,
$|\bar{I}_{nn}^{-}(\omega)|$. Over the range
$\omega_0=0.42$--$0.50$, the spectra nearly collapse onto a single
curve: the dominant peak remains at approximately the same energy,
and its line shape exhibits only weak variations. In particular, no
systematic peak splitting or avoided crossing is resolved as
$\omega_0$ is tuned through the graviton energy. The data therefore
provide no evidence for a simple coherent hybridization between an
isolated graviton mode and the bare phonon within the present
resolution.

We emphasize that each spectrum is normalized by its integrated
weight. This comparison consequently tests the graviton peak energy
and line shape, but does not determine the absolute spectral weight
in the negative-chirality channel or the magnitude of the chirality
deficit. As in the density-response calculations, varying $\omega_0$
at fixed $g$ also changes the dimensional electron--phonon vertex
$g\omega_0$.

\section*{Details on the perturbative treatment of Electron-phonon interaction in Landau Level}

In this section, we study the phonon-mediated contribution to the graviton chirality in a Landau level. The overall picture is that for scalar phonons coupling to the electron density, eliminating the phonon produces an effective interaction that can modify the electronic ground state and the corresponding graviton mode. We examine this mechanism for fully spin-polarized fermions at filling $\nu=1/3$ in the lowest Landau level (LLL). The continuum calculation isolates the effect of interaction renormalization. Note that it is not a microscopic reduction of the local Holstein model in the main text but provides a physical understanding of the chirality change in graviton modes. In the following discussion we set $\hbar=1$, so phonon frequencies have energy units.

\subsection*{Projected electron-phonon model}

The electron and its LLL projection are
\begin{equation}
 \hat{\rho}_{\bm q}=\sum_{i=1}^{N_e}\ee^{-i\bm q\cdot\hat{\bm r}_i},\qquad
 \hat{P}_{\mathrm{LLL}}\hat{\rho}_{\bm q}\hat{P}_{\mathrm{LLL}}
 =F_0(q)\hat{\bar{\rho}}_{\bm q},\qquad
 F_0(q)=\ee^{-q^2\ell_B^2/4},
 \label{smcg:density}
\end{equation}
where $q=|\bm q|$ and $\hat{\bar{\rho}}_{\bm q}$ is the guiding-center density. The baseline Hamiltonian is
\begin{equation}
 \hat{H}_{\mathrm{base}}=\frac{1}{2A}\sum_{\bm q\ne0}
 V_{\mathrm{base}}(q)F_0(q)^2
 \bigl(\hat{\bar{\rho}}_{\bm q}\hat{\bar{\rho}}_{-\bm q}-N_e\bigr).
 \label{smcg:Hbase}
\end{equation}
Here $A$ is the area and $V_{\mathrm{base}}(q)$ has units of energy times area. We use either the pure first-pseudopotential model or the Coulomb interaction. The LLL pseudopotentials are denoted by $v_m$ to distinguish them from the nearest-neighbor interaction $V_1$ in the lattice Hamiltonian. At fixed $N_e$, omitting the $\bm q=0$ term and subtracting the one-particle contribution changes only constants in the homogeneous continuum problem. Identity operators multiplying these constants are suppressed.

For a quasi-two-dimensional electron layer coupled to three-dimensional phonons, let $a=(j,\bm Q)$ label a mode, with $\bm Q=(\bm q,q_z)$ and $\epsilon_a=\omega_{j\bm Q}$. The projected coupling is\cite{smcgBenedict1999}
\begin{equation}
 \hat{H}_{\mathrm{ph}}=\sum_a\epsilon_a\hat{b}_a^\dagger\hat{b}_a,\qquad
 \hat{H}_{\mathrm{ep}}=\sum_a
 \bigl(\hat{G}_a\hat{b}_a+\hat{G}_a^\dagger\hat{b}_a^\dagger\bigr),\qquad
 \hat{G}_a=\frac{M_j(\bm Q)Z(q_z)}{\sqrt{\mathcal V}}
 F_0(q)\hat{\bar{\rho}}_{-\bm q}.
 \label{smcg:coupling}
\end{equation}
The confinement form factor is $Z(q_z)=\int dz\,|f(z)|^2\ee^{iq_zz}$, where $f(z)$ is the normalized perpendicular wavefunction. It is distinct from the LLL form factor $F_0(q)$. In-plane momentum is conserved. Perpendicular confinement is incorporated through $Z(q_z)$. For strictly two-dimensional phonons, the coupling becomes $\hat{G}_{\bm q}=g_qF_0(q)\hat{\bar{\rho}}_{-\bm q}/\sqrt A$. The amplitudes $M_j(\bm Q)$ and $g_q$ are dimensionful, unlike the Holstein parameter $g$ in the main text. We retain only scalar-density coupling and neglect phonon-assisted Landau-level mixing. A coupling that changes hopping or the orbital metric requires a different vertex.

\subsection*{Phonon elimination and the static approximation}

Although the uncoupled phonon vacuum has zero mean displacement, the coupling contributes through virtual processes of the form $\ket{n;0}\to\ket{m;1_a}\to\ket{n';0}$. For $\hat{H}_{\mathrm{base}}\ket n=E_n\ket n$, the intermediate energy cost is
$D_{mna}=E_m-E_n+\epsilon_a$. A Schrieffer-Wolff transformation eliminates the leading coupling between the phonon vacuum and its complement\cite{smcgBravyi2011}. With $\hat{H}_0=\hat{H}_{\mathrm{base}}+\hat{H}_{\mathrm{ph}}$ and
$\hat{P}_{\mathrm{vac}}=\hat{I}_{\mathrm e}\otimes\ket{0_{\mathrm{ph}}}\bra{0_{\mathrm{ph}}}$, its anti-Hermitian generator has first-order part
\begin{equation}
 \hat{S}_{\mathrm{SW}}^{(1)}=\sum_{n,m,a}
 \frac{\bra m\hat{G}_a^\dagger\ket n}{D_{mna}}
 \ket{m;1_a}\bra{n;0}-\mathrm{h.c.}
 \label{smcg:S}
\end{equation}
The Hermitian second-order correction is
\begin{equation}
 \bigl(\delta\hat{H}^{(2)}\bigr)_{n'n}
 =-\frac12\sum_{m,a}\bra{n'}\hat{G}_a\ket m\bra m\hat{G}_a^\dagger\ket n
 \left(\frac1{D_{mna}}+\frac1{D_{mn'a}}\right).
 \label{smcg:SW}
\end{equation}
The external states belong to a chosen low-energy window. The intermediate states need not lie in that window. Only one-phonon intermediate states contribute at this order, but no fixed phonon number is imposed on the microscopic model.

A finite-system diagnostic of weak virtual admixture is
\begin{equation}
 \eta_n^2=\sum_{m,a}
 \frac{|\bra m\hat{G}_a^\dagger\ket n|^2}{|D_{mna}|^2}\ll1.
 \label{smcg:admixture}
\end{equation}
The full mode sum must be small, not only its individual terms. Uniform control over a target subspace is a stronger requirement. In the thermodynamic limit, extensive vacuum dressing must be distinguished from the validity of a local effective expansion. A static reduction also requires
\begin{equation}
 \frac{|E_m-E_n|}{\epsilon_a}\ll1
 \quad\text{for appreciably coupled transitions from the target states}.
 \label{smcg:antiadiabatic}
\end{equation}
Weak coupling alone does not imply Eq.~\eqref{smcg:antiadiabatic}. A sufficient hierarchy for a gapped branch is
$\Delta_{\mathrm{rel}},k_BT\ll\Omega_0\ll\omega_{\mathrm{cyc}}$, together with weak coupling and the absence of inter-Landau-level resonance. Here $\Delta_{\mathrm{rel}}$ includes the connected electronic transitions. The cyclotron frequency $\omega_{\mathrm{cyc}}$ is distinct from the spectral cutoff $\omega_c$ used in the main text.

Approximating $D_{mna}$ by $\epsilon_a$ and summing over intermediate electronic states gives
\begin{equation}
 \hat{H}_{\mathrm{eff}}\simeq\hat{H}_{\mathrm{base}}
 -\sum_a\frac{\hat{G}_a\hat{G}_a^\dagger}{\epsilon_a},\qquad
 \delta V_{\mathrm{ph}}(q)
 =-2\sum_j\int\frac{dq_z}{2\pi}
 \frac{|M_j(\bm Q)Z(q_z)|^2}{\epsilon_a}.
 \label{smcg:staticV}
\end{equation}
Each $\hat{G}_a$ contains a density operator. The second-order process therefore generates a density product,
\begin{equation}
 \hat{\bar{\rho}}_{\bm q}\hat{\bar{\rho}}_{-\bm q}
 =N_e+\sum_{i\ne j}\ee^{-i\bm q\cdot\hat{\bm R}_i}
                         \ee^{i\bm q\cdot\hat{\bm R}_j},
 \label{smcg:pair}
\end{equation}
where $\hat{\bm R}_i$ is a guiding-center coordinate. The distinct-particle terms form the induced pair interaction. The factor of two in $\delta V_{\mathrm{ph}}$ follows from the $1/2$ convention in Eq.~\eqref{smcg:Hbase}. The effective Hamiltonian has the same form as Eq.~\eqref{smcg:Hbase}, with $V_{\mathrm{base}}$ replaced by $V_{\mathrm{base}}+\delta V_{\mathrm{ph}}$. The approximation neglects retardation. It does not replace the phonons by a prescribed static potential.

Integrating out harmonic phonons without this approximation gives the retarded kernel
\begin{equation}
 \delta V^R_{\mathrm{ph}}(q,E)=\sum_j\int\frac{dq_z}{2\pi}
 \frac{2\epsilon_a|M_j(\bm Q)Z(q_z)|^2}{(E+i0^+)^2-\epsilon_a^2},
 \label{smcg:retarded}
\end{equation}
where $E$ is the transferred energy. This elimination is exact within the projected harmonic model. Both LLL form factors remain outside the kernel. Evaluating Eq.~\eqref{smcg:retarded} at $E=0$ reproduces Eq.~\eqref{smcg:staticV}, but does not establish the scale separation in Eq.~\eqref{smcg:antiadiabatic}.

Without the static approximation, a second-order energy correction can be evaluated from the electronic resolvent,
\begin{equation}
 \Sigma_n(z)=\sum_a\bra n\hat{G}_a
 (z-\hat{H}_{\mathrm{base}}-\epsilon_a)^{-1}\hat{G}_a^\dagger\ket n
 =\sum_{m,a}\frac{|\bra m\hat{G}_a^\dagger\ket n|^2}{z-E_m-\epsilon_a}.
 \label{smcg:selfenergy}
\end{equation}
For an isolated graviton eigenstate $\ket G$, the excitation-energy shift is
$\operatorname{Re}\Sigma_G(E_G+i0^+)-\operatorname{Re}\Sigma_0(E_0+i0^+)$. A resonance $E_n=E_m+\epsilon_a$ requires a dynamical treatment. Equation~\eqref{smcg:selfenergy} is an alternative second-order calculation and must not be added to the full static correction.

\subsection*{Gaussian coupling and deformation of the Laughlin state}

The static elimination does not determine the momentum dependence of the interaction without specifying the coupling amplitude. We choose an isotropic Gaussian profile to vary the interaction strength and spatial range independently. This is a model assumption, not a Gaussian kernel derived from the acoustic coupling below. One realization is a dispersionless gapped branch with
\begin{equation}
 \omega_{\bm q,q_z}=\Omega_0,\qquad
 M(\bm q,q_z)=M_0h(q_z)\widetilde W_\xi(q),\qquad
 W_\xi(\bm r)=\frac{\ee^{-r^2/\xi_{\mathrm{ph}}^2}}{\pi\xi_{\mathrm{ph}}^2},\qquad
 \widetilde W_\xi(q)=\ee^{-q^2\xi_{\mathrm{ph}}^2/4}.
 \label{smcg:Gaussianvertex}
\end{equation}
For $\xi_{\mathrm{ph}}>0$, a scalar potential proportional to
$\int d^2r'\,W_\xi(\bm r-\bm r')\hat{u}(\bm r',z)$ samples the vibrational coordinate $\hat{u}$ over a finite region. The Gaussian is assumed in this spatial coupling. It is unrelated to the Gaussian LLL form factor. Substitution into Eq.~\eqref{smcg:staticV} yields
\begin{equation}
 C_z=\int\frac{dq_z}{2\pi}|h(q_z)Z(q_z)|^2,\qquad
 |g_0|^2=|M_0|^2C_z,\qquad
 \delta V_{\mathrm G}(q)=-U_{\mathrm G}\ee^{-q^2\xi_{\mathrm{ph}}^2/2},\qquad
 U_{\mathrm G}=\frac{2|g_0|^2}{\Omega_0}.
 \label{smcg:Gaussian}
\end{equation}
Two coupling amplitudes enter the virtual process, which doubles the exponent. For $h=1$ and
$|Z(q_z)|^2=(1+a_z^2q_z^2)^{-3}$, the confinement factor is $C_z=3/(16a_z)$ for $a_z>0$. In this separable construction, confinement changes the amplitude but not the in-plane profile. The equivalent two-dimensional coupling is $g_q=g_0\ee^{-q^2\xi_{\mathrm{ph}}^2/4}$. Its induced real-space pair interaction is
$\delta V_{\mathrm G}(r)=-U_{\mathrm G}\ee^{-r^2/(2\xi_{\mathrm{ph}}^2)}/(2\pi\xi_{\mathrm{ph}}^2)$.

Let $\beta=\xi_{\mathrm{ph}}^2/(2\ell_B^2)$ and
$\Lambda_{\mathrm G}=U_{\mathrm G}/(4\pi\ell_B^2)$. The amplitude $U_{\mathrm G}$ has units of energy times area, while $\Lambda_{\mathrm G}$ has units of energy. The planar LLL pseudopotential corrections are
\begin{equation}
 \delta v_m=\int_0^\infty\frac{q\,dq}{2\pi}\delta V_{\mathrm G}(q)
 L_m(q^2\ell_B^2)\ee^{-q^2\ell_B^2}
 =-\Lambda_{\mathrm G}\int_0^\infty dx\,\ee^{-(1+\beta)x}L_m(x)
 =-\Lambda_{\mathrm G}\frac{\beta^m}{(1+\beta)^{m+1}},
 \label{smcg:PP}
\end{equation}
where $x=q^2\ell_B^2$. For spin-polarized fermions only odd $m$ contribute, and
$\delta\hat{H}=\sum_{m\,\mathrm{odd}}\delta v_m\hat{\mathcal P}_m$, with
$\hat{\mathcal P}_m=\sum_{i<j}\hat{P}_m^{ij}$. At fixed $\Lambda_{\mathrm G}$, the magnitude $|\delta v_m|$ is maximal at $\beta=m$ for $m>0$. This does not determine the range that maximizes the chirality change.

Two limits separate energy renormalization from ground-state deformation. At $\beta=0$, only $\delta v_0=-\Lambda_{\mathrm G}$ survives, so the induced contact interaction is inactive for spin-polarized fermions. For $\beta\ll1$, $\delta v_1\sim-\Lambda_{\mathrm G}\beta$ and $\delta v_3\sim-\Lambda_{\mathrm G}\beta^3$. A change of $v_1$ alone in the pure model
$\hat{H}_{\mathrm{model}}=v_1\hat{\mathcal P}_1$ rescales the spectrum without changing its eigenvectors, provided $v_1+\delta v_1>0$. Changing $v_1$ in a Coulomb Hamiltonian is not an overall rescaling. Gap softening by itself therefore does not establish a loss of chirality.

For the model Laughlin state, a fixed circular quadrupolar probe satisfies \cite{liou2019chiral}
\begin{equation}
 \hat{O}_{\mathrm{opp}}^{(1)}\propto
 \sum_{i<j,M}\ket{3,M}_{ij}\bra{1,M}_{ij},\qquad
 \hat{O}_{\mathrm{opp}}^{(1)}\ket{\Psi_L}=0.
 \label{smcg:modelprobe}
\end{equation}
Here $M$ labels pair center-of-mass motion. The identity follows from the absence of $m=1$ pairs in $\ket{\Psi_L}$. It applies to this probe rather than to an arbitrary quadrupolar vertex. Its conjugate defines the active channel.

For $\hat{H}(\lambda)=\hat{H}_{\mathrm{model}}+\lambda\hat{W}$, with dimensionless $\lambda$, the ground state in a fixed sector is
\begin{equation}
 \ket{0(\lambda)}=\ket{\Psi_L}+\lambda\ket{\delta0}+O(\lambda^2),\qquad
 \ket{\delta0}=\sum_{r\notin\mathcal G_0}\ket r
 \frac{\bra r\hat{W}\ket{\Psi_L}}{E_0-E_r},
 \label{smcg:chi}
\end{equation}
where $\mathcal G_0$ is the unperturbed ground-state manifold. The ground-state deformation can activate the previously forbidden response. To leading order, its inelastic weight and the normalized $m=1$ pair amplitude obey
\begin{equation}
 W_{\mathrm{opp}}^{(1)}=
 \frac{\lambda^2}{N_e}\bigl\|(\hat{I}_{\mathrm e}-\hat{P}_{\mathcal G_0})
 \hat{O}_{\mathrm{opp}}^{(1)}\ket{\delta0}\bigr\|^2+O(\lambda^3),\qquad
 p_1=\frac{2\lambda^2}{N_e(N_e-1)}
 \bra{\delta0}\hat{\mathcal P}_1\ket{\delta0}+O(\lambda^3).
 \label{smcg:quadratic}
\end{equation}
The first state-changing term at short range is normally $\delta v_3$. Away from a closing gap, this gives
$W_{\mathrm{opp}}^{(1)}/W_{\mathrm{dom}}^{(1)}(0)
=O[(\Lambda_{\mathrm G}/v_1)^2\beta^6]$.
For a Coulomb baseline, the opposite-helicity amplitude can already be nonzero, so interference allows a linear change of either sign. The graviton energy is also not constrained to decrease by attraction alone. For an isolated level with $\Delta_G=E_G-E_0$,
\begin{equation}
 \left.\frac{d\Delta_G}{d\lambda}\right|_0
 =\sum_mw_m\bigl(\bra G\hat{\mathcal P}_m\ket G
                  -\bra0\hat{\mathcal P}_m\ket0\bigr),\qquad
 \hat{W}=\sum_mw_m\hat{\mathcal P}_m.
 \label{smcg:slope}
\end{equation}

\subsection*{Static acoustic piezoelectric interaction}

A separate interaction profile follows from acoustic piezoelectric coupling. We use the amplitudes of Benedict \textit{et al.}\cite{smcgBenedict1999}, who studied real phonon absorption, to calculate a zero-frequency interaction. We assume unscreened piezoelectric coupling with isotropic Debye dispersions. Perpendicular confinement is described by the Fang-Howard form factor. A uniform average over the in-plane azimuth removes the crystal's fourfold component. Thus
\begin{equation}
 |Z(q_z)|^2=(1+a_z^2q_z^2)^{-3},\qquad
 \omega_j(\bm Q)=c_jQ,\qquad Q^2=q^2+q_z^2.
 \label{smcg:FHpiezo}
\end{equation}
With the volume-independent coefficients
$\mathcal A_j=\mathcal V|\gamma_j^{\mathrm{pa}}|^2$, the averaged couplings are
\begin{equation}
 \overline{|M_L|^2}=\frac{\mathcal A_L}{Q}\frac{9q^4q_z^2}{8Q^6},\qquad
 \sum_{j\in T}\overline{|M_j|^2}
 =\frac{\mathcal A_T}{Q}\frac{q^2(q^4+8q_z^4)}{8Q^6}.
 \label{smcg:piezovertices}
\end{equation}
Both transverse polarizations are included in the second expression. Integrating over $q_z$ in Eq.~\eqref{smcg:staticV} gives
\begin{align}
 \delta V_{\mathrm{pz}}(q)&=-\frac{\mathcal K_{\mathrm{pz}}}{q}F(a_zq),\quad q>0,
 &\mathcal K_{\mathrm{pz}}&=\frac{1}{128}
 \left(\frac{9\mathcal A_L}{c_L}+\frac{13\mathcal A_T}{c_T}\right),
 \label{smcg:pzkernel}\\
 F(y)&=w_LF_L(y)+(1-w_L)F_T(y),
 &w_L&=\frac{9/c_L^2}{9/c_L^2+13/c_T^2},
 \label{smcg:pzweights}\\
 F_L(y)&=\frac{1+6y+12y^2+2y^3}{(1+y)^6},
 &F_T(y)&=\frac{13+78y+72y^2+82y^3+36y^4+6y^5}{13(1+y)^6}.
 \label{smcg:pzformfactors}
\end{align}
The common material prefactor gives $\mathcal A_j\propto c_j^{-1}$. The speed ratio $c_L/c_T=0.0305/0.0196$ used in Ref.~\cite{smcgBenedict1999} yields $w_L\simeq0.2223$. No nonequilibrium phonon occupation enters this kernel. In the same ideal static limit, deformation-potential coupling produces the momentum-independent term
$-3\Xi_0^2/(16\rho_Mc_L^2a_z)$ for $a_z>0$. Here $\Xi_0$ is the deformation potential and $\rho_M$ is the mass density. This term contributes only $v_0$ and is inactive for spin-polarized fermions.

The confinement parameter $\alpha=a_z/\ell_B$ is distinct from the in-plane Gaussian range $\beta$. Since $F(0)=1$, the zero-thickness correction is proportional to $-1/q$. Against a strict Coulomb baseline it only rescales the Hamiltonian while the net coefficient remains positive. Finite $\alpha$ changes the pseudopotential ratios. Quantitative material modeling would require the same confinement in the bare Coulomb interaction. Because acoustic phonons are gapless, Eq.~\eqref{smcg:pzkernel} is used here as a static interaction benchmark. Its dynamical validity requires a separate mode-resolved check of Eq.~\eqref{smcg:antiadiabatic}.

\subsection*{Circular response and ground-state correlations}

A momentum-space realization of the fixed model probe is
\begin{equation}
 \hat{O}_\pm^{(1)}=\frac{\ell_B^2}{2A}\sum_{\bm q\ne0}
 (q_x\pm iq_y)^2V^{\mathrm{ref}}(q)F_0(q)^2
 \bigl(\hat{\bar{\rho}}_{\bm q}\hat{\bar{\rho}}_{-\bm q}-N_e\bigr),\qquad
 V^{\mathrm{ref}}(q)=4\pi\ell_B^2v_1L_1(q^2\ell_B^2).
 \label{smcg:Ofixed}
\end{equation}
The factor $\ell_B^2$ gives the operator energy units without changing its circular-weight ratio. Keeping this probe fixed isolates the effect of changing the electronic state. It need not coincide with the microscopic experimental probe. It also differs from the unprojected lattice density operator used in the main text.

For a specified probe pair $\hat{X}_\pm$, define the LLL inelastic spectral density as
\begin{equation}
 \begin{aligned}
 I_\pm^X(E)&=\frac1{N_e}\sum_{n\notin\mathcal G}
 |\bra n\hat{X}_\pm\ket0|^2\delta\bigl(E-(E_n-E_0)\bigr),\\
 W_\pm^X&=\int_0^\infty dE\,I_\pm^X(E)
 =\frac1{N_e}\bra0\hat{X}_\pm^\dagger
 (\hat{I}_{\mathrm e}-\hat{P}_{\mathcal G})\hat{X}_\pm\ket0,
 \end{aligned}
 \label{smcg:spectrum}
\end{equation}
where $\mathcal G$ is the ground-state manifold of $\hat{H}_{\mathrm{eff}}$. Completeness also gives the total probe-generated norm,
\begin{equation}
 \frac{\|\hat{X}_\pm\ket0\|^2}{N_e}
 =\frac{\bra0\hat{X}_\pm^\dagger\hat{X}_\pm\ket0}{N_e}
 =W_\pm^X+\frac{\|\hat{P}_{\mathcal G}\hat{X}_\pm\ket0\|^2}{N_e}.
 \label{smcg:groundstateweights}
\end{equation}
For a fixed probe, these total weights are equal-time ground-state correlations. The inelastic weights additionally exclude transitions within the ground-state manifold. This is the ground-state dependence of the circular response used in the interaction mechanism. The energy-resolved spectrum, including a ratio restricted to a graviton peak, also depends on the excited states.

We fix the minus label as the reference dominant helicity and define
$\mathcal C_{\mathrm{inel}}^X=(W_-^X-W_+^X)/(W_-^X+W_+^X)$. The sign is not relabeled when the other channel becomes dominant. Independent normalization of the two channels removes their relative weight and is suitable only for displaying spectral shapes.

The full stress of the static interaction is a different probe\cite{nguyen2022multiple}. Writing
$K(q)=V_{\mathrm{tot}}(q)\ee^{-q^2\ell_B^2/2}$, with
$V_{\mathrm{tot}}=V_{\mathrm{base}}+\delta V_{\mathrm{ph}}$, gives
\begin{equation}
 \hat{T}_\pm=\frac{1}{8A}\sum_{\bm q\ne0}
 \frac{(q_x\pm iq_y)^2}{q}\frac{dK(q)}{dq}
 \bigl(\hat{\bar{\rho}}_{\bm q}\hat{\bar{\rho}}_{-\bm q}-N_e\bigr).
 \label{smcg:stress}
\end{equation}
The derivative acts on the interaction as well as the LLL form factor. Substituting $V_{\mathrm{tot}}$ for $V^{\mathrm{ref}}$ in Eq.~\eqref{smcg:Ofixed} does not give Eq.~\eqref{smcg:stress}. A parameter-dependent stress response includes changes in the probe itself, in addition to ground-state deformation.

An isotropic planar interaction can activate the opposite-helicity response without directly mixing angular momenta $+2$ and $-2$. On a torus this distinction depends on the rotation group of the finite cluster. The $C_6$ representations distinguish these helicities, whereas $C_4$ does not. A hexagonal momentum plotting cell alone does not establish $C_6$ symmetry. On a sphere, the $M=\pm2$ components of a single $L=2$ multiplet are not the two planar helicities.

\subsection*{Exact-diagonalization results}

We diagonalize the static LLL Hamiltonian for $N_e=8$ electrons at $\nu=1/3$. The momentum labels use $N_x=4$ and $N_y=6$, giving $N_\phi=N_xN_y=24$. The Gaussian strength is denoted by $s_{\mathrm G}$ and the piezoelectric strength by $s_{\mathrm{pz}}$. Both are labeled $U_{\mathrm{ph}}$ in the figures. Each strength is scanned from $0.5$ to $5$ in steps of $0.5$. The same grid is used for $\beta$ or $\alpha$. These scan parameters describe effective interactions and are not calibrated material couplings. They are not identified with the dimensional amplitude $U_{\mathrm G}$. A quantitative comparison between baselines requires a common Hamiltonian normalization.

The maps use the numerical probe-generated norms,
\begin{equation}
 \mathcal N_\pm=\bigl\|\hat{O}_\pm^{\mathrm{num}}\ket0\bigr\|^2,\qquad
 \mathcal C=\frac{\mathcal N_--\mathcal N_+}{\mathcal N_-+\mathcal N_+},\qquad
 1-\mathcal C=\frac{2\mathcal N_+}{\mathcal N_-+\mathcal N_+}.
 \label{smcg:seed}
\end{equation}
A purely negative-helicity response has $\mathcal C=1$. The chirality deficit $1-\mathcal C$ is shown on a logarithmic color scale. The ratio is undefined when both norms vanish. As written, Eq.~\eqref{smcg:seed} includes any ground-manifold contribution. Its identification with $\mathcal C_{\mathrm{inel}}^X$ requires the same probe and a vanishing or removed ground-manifold contribution. We retain $\hat{O}_\pm^{\mathrm{num}}$ separately from the analytic reference probe because their identification requires the numerical operator convention.

\textit{Gaussian interaction.}
Figure~\ref{smcg:fig-gaussian-v1} shows the Gaussian correction to the $v_1$ model. The chirality deficit remains small over a broad region, but increases with strength along much of the Laughlin-like regime. At $(s_{\mathrm G},\beta)=(2,4)$, the three-state low-energy manifold remains separated from higher levels. The circular response is predominantly in the $\chi=-1$ channel, although $\mathcal C<1$. Thus the norm-based response can depart from its model-state chirality before the characteristic low-energy structure is lost.

At $(s_{\mathrm G},\beta)=(4,2)$, the separated three-state structure is replaced by a reorganized set of low-energy levels. The structure factor is enhanced at two opposite nonzero momenta. These observations indicate the breakdown of the Laughlin-like regime in the finite system and are consistent with competing charge order, possibly of stripe character. The available size does not determine a thermodynamic ordered phase. The large chirality deficit in this region accompanies a change of state rather than a weakened graviton within the original liquid.

\begin{figure*}[t]
 \centering
 \includegraphics[width=\linewidth]{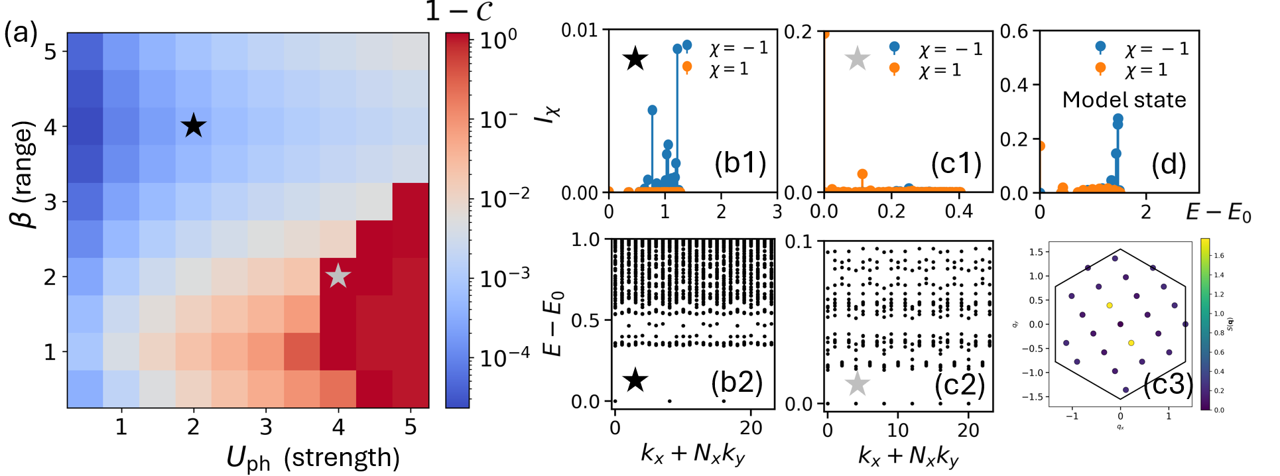}
 \caption{\textbf{Gaussian correction to the LLL $v_1$ model.}
 Results for $N_e=8$ and $N_\phi=24$, with $N_x=4$ and $N_y=6$.
 \textbf{(a)} Chirality deficit $1-\mathcal C$ from Eq.~\eqref{smcg:seed} on a logarithmic color scale.
 The horizontal label $U_{\mathrm{ph}}$ denotes $s_{\mathrm G}$, while $\beta=\xi_{\mathrm{ph}}^2/(2\ell_B^2)$.
 The black and gray stars mark $(s_{\mathrm G},\beta)=(2,4)$ and $(4,2)$.
 \textbf{(b1), (c1)} Circular response spectra at the corresponding points.
 Blue and orange stems denote $\chi=-1$ and $\chi=+1$.
 \textbf{(b2), (c2)} Low-energy many-body spectra resolved by $k_x+N_xk_y$, measured relative to $E_0$.
 \textbf{(c3)} Static structure factor at the gray-star point. Enhanced correlations at opposite nonzero momenta are consistent with competing charge order.
 \textbf{(d)} Circular response of the reference Laughlin model state. Perfect chirality of the fixed model probe follows from Eq.~\eqref{smcg:modelprobe}. Separately normalized spectral shapes do not determine the relative channel weights.}
 \label{smcg:fig-gaussian-v1}
\end{figure*}

The Coulomb results in Fig.~\ref{smcg:fig-gaussian-coulomb} show a larger departure from perfect chirality over the displayed scan. At $(s_{\mathrm G},\beta)=(2,4)$, the three-state manifold remains separated from the excited levels, while the opposite-helicity contribution is appreciable in the norm-based measure. At $(4,2)$, the low-energy spectrum is reorganized and the negative-helicity response is strongly suppressed. This comparison concerns the displayed scans rather than matched microscopic couplings or baseline energy scales.

\begin{figure*}[t]
 \centering
 \includegraphics[width=\linewidth]{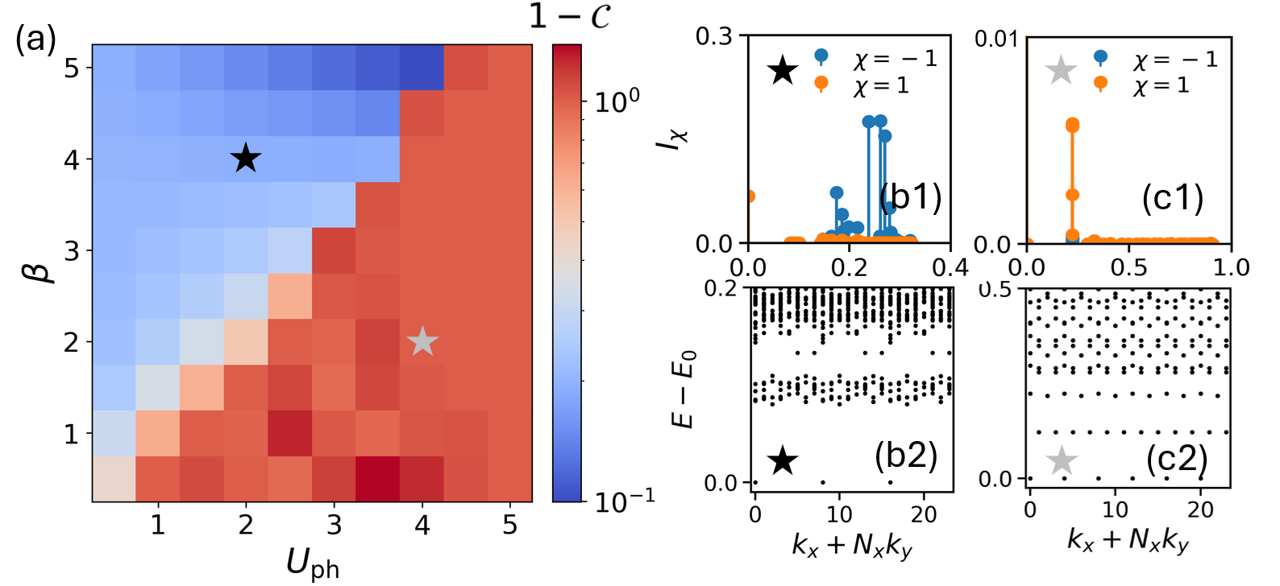}
 \caption{\textbf{Gaussian correction to the Coulomb interaction.}
 The system size is the same as in Fig.~\ref{smcg:fig-gaussian-v1}.
 \textbf{(a)} Chirality deficit $1-\mathcal C$ versus $s_{\mathrm G}$, labeled $U_{\mathrm{ph}}$, and $\beta$.
 The color scale is logarithmic and has different limits from Fig.~\ref{smcg:fig-gaussian-v1}.
 The black and gray stars indicate $(s_{\mathrm G},\beta)=(2,4)$ and $(4,2)$.
 \textbf{(b1), (c1)} Circular response spectra at the marked points, plotted against $E-E_0$.
 Blue and orange stems represent $\chi=-1$ and $\chi=+1$.
 \textbf{(b2), (c2)} Corresponding many-body spectra.
 The separated three-state manifold in \textbf{(b2)} is replaced by a reorganized low-energy structure in \textbf{(c2)}.}
 \label{smcg:fig-gaussian-coulomb}
\end{figure*}

\textit{Piezoelectric interaction.}
For the $v_1$ baseline, Fig.~\ref{smcg:fig-piezo-v1} shows a similar evolution under the static piezoelectric correction. At fixed $\alpha$, increasing the strength raises the chirality deficit over much of the scan. The point $(s_{\mathrm{pz}},\alpha)=(2,4)$ retains a separated three-state manifold and a finite-energy response dominated by $\chi=-1$. At $(4,2)$, the low-energy spectrum is qualitatively different and the $\chi=+1$ response becomes prominent. Increasing $\alpha$ suppresses the correction through $F(\alpha q\ell_B)$, moving the region of large changes toward stronger coupling.

\begin{figure*}[t]
 \centering
 \includegraphics[width=\linewidth]{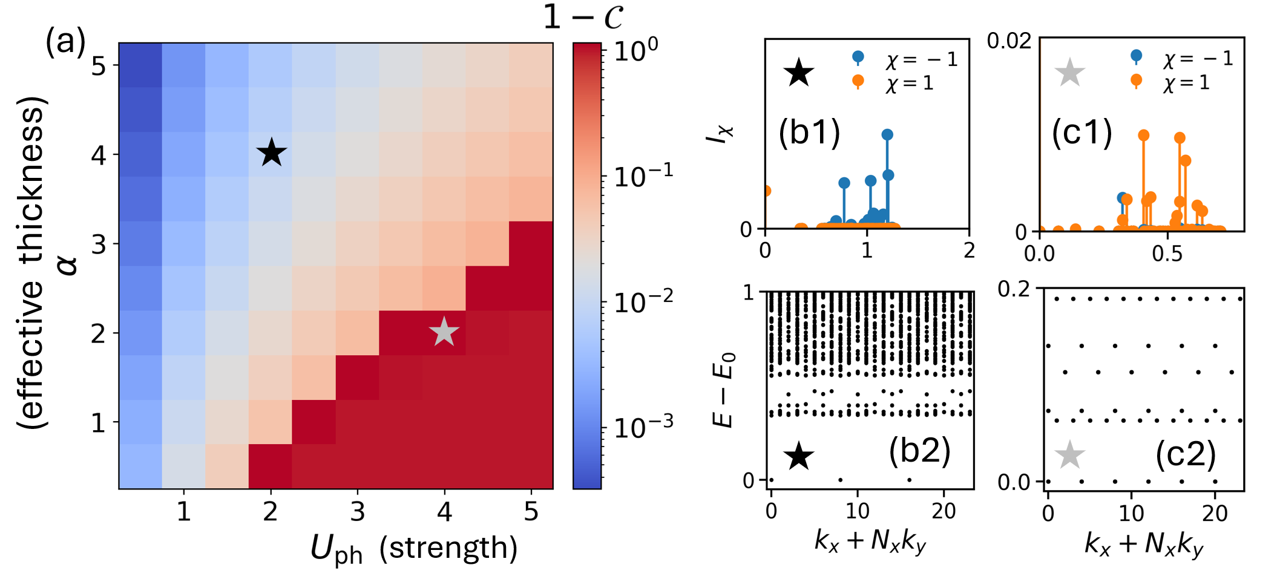}
 \caption{\textbf{Static piezoelectric correction to the LLL $v_1$ model.}
 Results for $N_e=8$ and $N_\phi=24$.
 \textbf{(a)} Chirality deficit $1-\mathcal C$ on a logarithmic color scale versus $s_{\mathrm{pz}}$, labeled $U_{\mathrm{ph}}$, and $\alpha=a_z/\ell_B$.
 The black and gray stars mark $(s_{\mathrm{pz}},\alpha)=(2,4)$ and $(4,2)$.
 \textbf{(b1), (c1)} Circular response spectra at these points, plotted against $E-E_0$.
 Blue and orange stems denote $\chi=-1$ and $\chi=+1$.
 \textbf{(b2), (c2)} Corresponding many-body spectra resolved by $k_x+N_xk_y$.
 The change in circular response is accompanied by a reorganization of the low-energy levels.}
 \label{smcg:fig-piezo-v1}
\end{figure*}

The Coulomb response in Fig.~\ref{smcg:fig-piezo-coulomb} need not vary monotonically. At some values of $\alpha$, increasing the attraction initially reduces $1-\mathcal C$. This improvement reverses at larger strength. It is consistent with the sign-indefinite interference allowed around a non-parent baseline. Stronger coupling produces a substantial loss of the original chirality, accompanied by a low-energy spectrum that no longer resembles the Laughlin-like manifold.

\begin{figure*}[t]
 \centering
 \includegraphics[width=\linewidth]{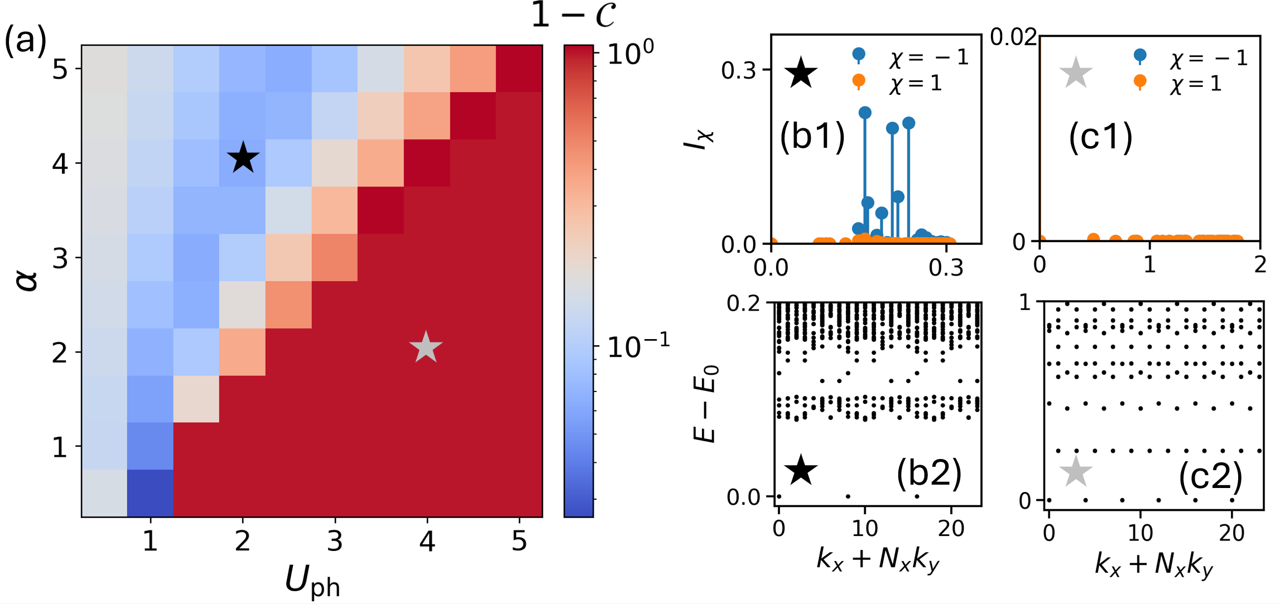}
 \caption{\textbf{Static piezoelectric correction to the Coulomb interaction.}
 The system size is the same as in Fig.~\ref{smcg:fig-gaussian-v1}.
 \textbf{(a)} Chirality deficit $1-\mathcal C$ versus $s_{\mathrm{pz}}$, labeled $U_{\mathrm{ph}}$, and $\alpha$, on a logarithmic color scale.
 Over part of the scan, increasing the attraction first reduces the deficit before increasing it at stronger coupling.
 The black and gray stars denote $(s_{\mathrm{pz}},\alpha)=(2,4)$ and $(4,2)$.
 \textbf{(b1), (c1)} Circular response spectra at the marked points, plotted against $E-E_0$.
 Blue and orange stems represent $\chi=-1$ and $\chi=+1$.
 \textbf{(b2), (c2)} Corresponding low-energy many-body spectra resolved by $k_x+N_xk_y$.}
 \label{smcg:fig-piezo-coulomb}
\end{figure*}

These results support an interaction-mediated reduction of the circular-response chirality while the low-energy structure remains Laughlin-like. At stronger coupling, the induced interaction destabilizes that regime. The numerical conclusion concerns the norm-based diagnostic in Eq.~\eqref{smcg:seed}. A quantitative statement about inelastic graviton chirality requires the probe identification and ground-manifold subtraction described above. The finite-size scans do not establish a thermodynamic phase boundary or the acoustic frequency hierarchy. They nevertheless show that phonon-mediated interaction changes can weaken the chiral response and ultimately destroy the Laughlin-like ground-state regime.

\subsection*{Relation to lattice phonons and acoustic probes}

The connection to a microscopic electron-phonon spectrum also requires transforming the observable. For a purely electronic probe,
\begin{equation}
 \hat{O}_{\mathrm{eff}}
 =\hat{P}_{\mathrm{vac}}\ee^{\hat{S}_{\mathrm{SW}}}\hat{O}
  \ee^{-\hat{S}_{\mathrm{SW}}}\hat{P}_{\mathrm{vac}}
 =\hat{P}_{\mathrm{vac}}\hat{O}\hat{P}_{\mathrm{vac}}
 +\frac12\hat{P}_{\mathrm{vac}}
 [\hat{S}_{\mathrm{SW}}^{(1)},[\hat{S}_{\mathrm{SW}}^{(1)},\hat{O}]]
 \hat{P}_{\mathrm{vac}}+\cdots.
 \label{smcg:probeSW}
\end{equation}
The first-order vacuum-block term vanishes. Probe dressing can contribute when the leading response is forbidden. Since $\Lambda_{\mathrm G}\propto|g_0|^2$, the weak-deformation static model-probe leakage is normally fourth order in the microscopic coupling amplitude. One-phonon sidebands in the full spectrum can carry second-order weight. The contact control and the model-state scaling do not describe the entire finite-frequency electron-phonon response.

The local spinless Holstein coupling provides an important distinction. Using the convention of the main text,
\begin{equation}
 \hat{H}_{\mathrm{ep}}^{\mathrm{lat}}
 =g\omega_0\sum_i(\hat{n}_i-\bar n)(\hat{b}_i+\hat{b}_i^\dagger),\qquad
 \delta\hat{H}_{\mathrm{loc}}
 =-g^2\omega_0\sum_i(\hat{n}_i-\bar n)^2.
 \label{smcg:Holstein}
\end{equation}
For one spinless orbital per site, $\hat{n}_i^2=\hat{n}_i$. At fixed filling the last sum equals $(1-2\bar n)N_e+N_s\bar n^2$, where $N_s$ is the number of sites, and is therefore constant. The leading local instantaneous term does not reshape pair interactions. Nontrivial lattice effects involve phonon dressing of hopping or finite-frequency processes. A nonlocal coupling can generate a different static interaction. The continuum calculations above isolate such an interaction mechanism, not an instantaneous Gaussian replacement for the local Holstein model. Electronic correlations in the static model are treated by exact diagonalization. The approximation concerns the elimination of phonons.

The acoustic proposal of Yang in Ref.~\cite{yang2016acoustic} concerns a different coupling channel. An imposed coherent strain changes the orbital metric and can resonantly excite the graviton while remaining slow compared with cyclotron motion. Our calculation instead treats equilibrium virtual exchange through scalar density. A uniform scalar potential couples only to the conserved particle number, whereas a uniform metric deformation remains active. The two descriptions are complementary. Equilibrium interaction renormalization does not make acoustic graviton spectroscopy unsuitable, and the resonant probe cannot be eliminated with the static approximation used here.

\end{widetext}
\end{document}